# A New Model-Free Method for MIMO Systems and Discussion on Model-Free or Model-Based

Feilong Zhang

*Abstract*—**Current model-free adaptive control (MFAC) can hardly deal with the time delay problem in multiple-input multiple-output (MIMO) systems. To solve this problem, a novel model-free adaptive predictive control (MFAPC) method is proposed. Compared to the current MFAC, i) the proposed method is based on a kind of prediction model which derives from the equivalent-dynamic-linearization model (EDLM); ii) the previous assumptions are relaxed and the application range of MFAPC are extended. The leading coefficient of the control input vector in system description is no more restricted to the diagonally dominant square matrix and the permissible ranges of pseudo orders $L_y$ and $L_u$ are extended; iii) the performance analysis and the issue of how to choose the matrix $\lambda$ are completed by an easy manner of analyzing the function of the closed-loop poles, however, both problems may not be realized by the previous contraction mapping method.**

## I. INTRODUCTION

The topic of MFAC has drawn considerable attention recently. The conception of model-free is under the claim that the controller design only relies on the I/O data of systems. Since the pseudo-gradient (PG) vector or pseudo-Jacobian matrix (PJM), whose elements act as the coefficients of the EDLM and the controller, can be estimated online. One merit of this kind of method is its focus on the discrete-time situation, aligning with the requirements of computer systems. MFAC has been extensively tested in several fields, such as process control, motor control, power systems and microgrids [1]-[21]. However, many of them contradict with [22]-[25].

Compared to the SISO system, fewer articles discuss MFAC in multivariable systems. Due to the presence of cross-coupling among input and output variables and the non-commutative nature of matrix multiplication, the conclusions drawn from SISO systems cannot be directly applied to nontrivial MIMO systems. Additionally, the time delay problem becomes more challenging. In [26], it is crucial to determine the delay structure described by the non-diagonal interactor matrix through a preliminary experiment. Failure to obtain or under/overestimate the time delays, or if they are time-varying, can result in poor control performance and even system instability [27].

Manuscript received Dec 3, 2020. This work was supported in part by the xxxxxxxxxx.

Feilong Zhang is with the State Key Laboratory of Robotics, Shenyang Institute of Automation, Chinese Academy of Sciences, Shenyang 110016, China (e-mail: zhangfeiong@sia.cn).

Furthermore, [28] has representively applied the interactor matrix in almost every adaptive control method, and a lot of pseudo exchange matrixes (PEM) are necessary for the design and performance analysis of GMVC and STC. However, this approach leads to a more complex controller design and poses greater challenges in practical applications. On the other hand, much literature focuses on decoupling controllers. However, the effectiveness of decoupling is relevant to the precision of the modeling [29], [30]. In the author's opinion, the effect of a decoupling controller may seldom be better than that of a directly designed controller in [28] when the estimated model is not precise enough. Besides, decoupling control is based on the condition that the number of outputs is equal to the number of inputs. In reality, the number of control variables may be greater than the number of system outputs, which can help achieve better performance.

The concept of model-free presented for nonlinear systems in [1]-[21] is not straightforward for operating engineers to understand properly and to master its essence. In this paper, we analyze this class of controller in linear deterministic finite-dimensional systems in simulations to exhibit its working principle more clearly, whether the controller aims to the known linear or unknown nonlinear systems. Because the fundamental tool of the current MFAC is to describe the nonlinear system model by the EDLM at each time based on the definition of differentiability, the controller design is inherently based on the linear model. The adaptability to a nonlinear system or uncertainty would be achieved by combining one online parameter estimation with any control law, according to the certainty equivalence adaptive control [26], [27].

In light of the discussion above and motivated by the easier implementation of the controller in industrial settings, we propose the MIMO-MFAPC based on the prediction model derived from the full-form EDLM.

The main contributions of this work are summarized as follows:

1) We extend the full-form EDLM and present its poof in Appendix. This novel process model offers a more objective reflection of the real plant, as it allows the coefficients of the model to be unrestricted by the square matrix. Additionally, the pseudo orders are not limited to the range of $1\le L_y\le n_y$ and $1\le L_u\le n_u$. Actually, the ideal choice is $L_y=n_y+1$ and $L_u=n_u+1$ with the purpose of a pseudo-Jacobin matrix acting as the real Jacobin matrix.

2) On the basis of EDLM, a prediction model with $N$-step is proposed as the fundamental tool of the MIMO-MFAPC design which is the optimal solution of the quadratic performance index. The determination of the interactor matrix, which determines the time delays between input and output variables, is unnecessary in the controller design process. It is required to set the prediction step $N$ larger than the maximum time delay among input variables and output variables since MFAPC generally deals with the time delay problem when the online estimated coefficients of EDLM are able to reflect the system objectively, even when the time delays are time-varying. By comparison, the current MFAC is restricted by a harsh assumption that the leading coefficient of the control input vector in the system description is a special kind of diagonally dominant square matrix. Additionally, any time delay might cause the invalidation of the current MFAC, even though all the system parameters are set to the actual values when the controlled system is linear deterministic finite-dimensional in simulation.

3) It is crucial to have a profound discussion, not only to provide the guidelines for controller design and performance analysis but also to identify the pitfalls and limitations of this type of adaptive controller. These are related to the success or failure of practical applications. To this end, i) the selection guide of the critical parameter matrix $\boldsymbol{\lambda}$ should be determined through quantitative performance analysis, which is distinct from the incorrectly proven conclusion that $\boldsymbol{\lambda}$ should be sufficiently large to guarantee the convergence of tracking error in [1], [2], [21]. ii) the essence of "model-free" or "model-based" is discussed in *Remark 1*. This may help us know why and how to implement the proposed method.

The rest of the paper is organized as follows. Section II presents the corrected EDLM for a class of discrete-time MIMO nonlinear systems. In Section III, the MFAPC design and its performance analysis are presented. Section IV presents the comparison results of simulations to show the proposed method's advantages and lead to further discussions. The conclusion is given in Section V. At last, Appendix I presents the proof of corrected EDLM, and Appendix II supplies some coefficient matrices for the iterative controller.

## II. Description of System and Prediction Model

### A. System Description

The discrete-time MIMO nonlinear system is considered as:

$$\begin{aligned} &\boldsymbol{y}(k+1)=\boldsymbol{f}(\boldsymbol{\varphi}(k)) \\ &\boldsymbol{\varphi}(k)=[\boldsymbol{y}(k),\cdots,\boldsymbol{y}(k-n_y),\boldsymbol{u}(k),\cdots,\boldsymbol{u}(k-n_u)] \end{aligned} \tag{1}$$

where $\boldsymbol{f}(\cdots)=\left[f_1(\cdots),\cdots,f_{My}(\cdots)\right]^T$ is the nonlinear vector-valued function, $n_y+1$, $n_u+1 \in$ Z are the orders of output vector $\boldsymbol{y}(k)$, input vector $\boldsymbol{u}(k)$ of the system at time $k$, respectively. The dimension of $\boldsymbol{y}(k)$ is $M_y$ and the dimension of $\boldsymbol{u}(k)$ is $M_u$.

*Assumption 1:* The partial derivatives of $\boldsymbol{f}(\cdots)$ with respect to all variables are continuous.

*Theorem 1*: If the system (1) satisfies the above assumption and $\Delta\boldsymbol{H}(k)\neq\boldsymbol{0}$, $0\le L_y$, $1\le L_u$, there must exist a pseudo-Jacobian matrix $\boldsymbol{\phi}_L^T(k)$ and (1) can be transformed into:

$$\Delta\boldsymbol{y}(k+1)=\boldsymbol{\phi}_L^T(k)\Delta\boldsymbol{H}(k) \tag{2}$$

where

$\boldsymbol{\phi}_L^T(k)=[\boldsymbol{\phi}_{Ly}^T(k),\boldsymbol{\phi}_{Lu}^T(k)]$, $\boldsymbol{\phi}_{Ly}^T(k)=[\boldsymbol{\Phi}_1(k),\cdots,\boldsymbol{\Phi}_{Ly}(k)]_{My\times(Ly\cdot My)}$,

$\boldsymbol{\phi}_{Lu}^T(k)=[\boldsymbol{\Phi}_{Ly+1}(k),\cdots,\boldsymbol{\Phi}_{Ly+Lu}(k)]_{My\times(Lu\cdot Mu)}$,

$\boldsymbol{\Phi}_i(k)\in\boldsymbol{R}^{m\times n}$: $m\times n=M_y\times M_y$, $i$=1,⋯, $L_y$; $m\times n=M_y\times M_u$, $i$=$L_y$+1,⋯, $L_y$+$L_u$; Define $\boldsymbol{H}(k)=\left[\boldsymbol{Y}_{Ly}^T(k)\quad \boldsymbol{U}_{Lu}^T(k)\right]^T$ and then $\Delta\boldsymbol{H}(k)=\left[\Delta\boldsymbol{Y}_{Ly}^T(k)\quad \Delta\boldsymbol{U}_{Lu}^T(k)\right]^T$ is a vector that contains the increment of system output vector $\Delta\boldsymbol{Y}_{Ly}(k)=[\Delta\boldsymbol{y}^T(k),\cdots,\Delta\boldsymbol{y}^T(k-L_y+1)]^T$ and the increment of control input vector $\Delta\boldsymbol{U}_{Lu}(k)=[\Delta\boldsymbol{u}^T(k),\cdots,\Delta\boldsymbol{u}^T(k-L_u+1)]^T$ within $[k-L_y+1,k]$ and $[k-L_u+1,k]$. The integers $L_y$ $(0\le L_y)$ and $L_u$ $(1\le L_u)$ are the pseudo orders.

$\boldsymbol{d}=\begin{bmatrix} d_{11} & \cdots & d_{1Mu} \\ \vdots & \ddots & \vdots \\ d_{My1} & \cdots & d_{MyMu} \end{bmatrix}$ is assumed to be the time delay matrix between the system input and output ($d_{ij}$=the time delay between the $j$-th input variable and the $i$-th output variable).

### B. Prediction Model

We can rewrite (2) into (3).

$$\boldsymbol{y}(k+1)=\boldsymbol{y}(k)+\boldsymbol{\phi}_L^T(k)\Delta\boldsymbol{H}(k) \tag{3}$$

Herein, we define

$$\boldsymbol{A}=\begin{bmatrix} \boldsymbol{0} & & & \\ \boldsymbol{I} & \boldsymbol{0} & & \\ & \ddots & \ddots & \\ & & \boldsymbol{I} & \boldsymbol{0} \end{bmatrix}_{(L_u\cdot M_u)\times(L_u\cdot M_u)} \qquad \boldsymbol{C}=\begin{bmatrix} \boldsymbol{0} & & & \\ \boldsymbol{I} & \boldsymbol{0} & & \\ & \ddots & \ddots & \\ & & \boldsymbol{I} & \boldsymbol{0} \end{bmatrix}_{(L_y\cdot M_y)\times(L_y\cdot M_y)}$$

$$\boldsymbol{B}^T=\begin{bmatrix}\boldsymbol{I} & \boldsymbol{0} & \cdots & \boldsymbol{0}\end{bmatrix}_{M_u\times(L_u\cdot M_u)} \qquad \boldsymbol{D}^T=\begin{bmatrix}\boldsymbol{I} & \boldsymbol{0} & \cdots & \boldsymbol{0}\end{bmatrix}_{M_y\times(L_y\cdot M_y)}$$

Define $\boldsymbol{A}^i=\boldsymbol{0}$ and $\boldsymbol{C}^i=\boldsymbol{0}$, $i=-1,-2,\cdots$ for the following description. The finite $N$-step forward prediction model is given as (4).

$$\begin{aligned} \Delta\boldsymbol{y}(k+1)&=\boldsymbol{\phi}_L^T(k)\Delta\boldsymbol{H}(k) \\ &=\boldsymbol{\phi}_{Ly}^T(k)\Delta\boldsymbol{Y}_{Ly}(k)+\boldsymbol{\phi}_{Lu}^T(k)\boldsymbol{A}\Delta\boldsymbol{U}_{Lu}(k-1)+\boldsymbol{\phi}_{Lu}^T(k)\boldsymbol{B}\Delta\boldsymbol{u}(k) \\ \Delta\boldsymbol{y}(k+2)&=\boldsymbol{\phi}_{Ly}^T(k)\Delta\boldsymbol{Y}_{Ly}(k+1)+\boldsymbol{\phi}_{Lu}^T(k)\boldsymbol{A}\Delta\boldsymbol{U}_{Lu}(k) \\ &\quad+\boldsymbol{\phi}_{Lu}^T(k)\boldsymbol{B}\Delta\boldsymbol{u}(k+1) \\ \Delta\boldsymbol{y}(k+3)&=\boldsymbol{\phi}_{Ly}^T(k)\Delta\boldsymbol{Y}_{Ly}(k+2)+\boldsymbol{\phi}_{Lu}^T(k)\boldsymbol{A}\Delta\boldsymbol{U}_{Lu}(k+1) \\ &\quad+\boldsymbol{\phi}_{Lu}^T(k)\boldsymbol{B}\Delta\boldsymbol{u}(k+2) \\ &\ \ \vdots \\ \Delta\boldsymbol{y}(k+N)&=\boldsymbol{\phi}_{Ly}^T(k)\boldsymbol{C}^{N-1}\Delta\boldsymbol{Y}_{Ly}(k)+\boldsymbol{\phi}_{Lu}^T(k)\boldsymbol{A}^N\Delta\boldsymbol{U}_{Lu}(k-1) \\ &\quad+\boldsymbol{\phi}_{Ly}^T(k)\boldsymbol{C}^{N-2}\boldsymbol{D}\Delta\boldsymbol{y}(k+1)+\cdots \\ &\quad+\boldsymbol{\phi}_{Ly}^T(k)\boldsymbol{D}\Delta\boldsymbol{y}(k+N-1)+\boldsymbol{\phi}_{Lu}^T(k)\boldsymbol{A}^{N-1}\boldsymbol{B}\Delta\boldsymbol{u}(k) \\ &\quad+\boldsymbol{\phi}_{Lu}^T(k)\boldsymbol{A}^{N-2}\boldsymbol{B}\Delta\boldsymbol{u}(k+1)+\cdots \\ &\quad+\boldsymbol{\phi}_{Lu}^T(k)\boldsymbol{B}\boldsymbol{u}(k+N-1) \end{aligned} \tag{4}$$

where $N$ denotes the predictive step length, $\Delta\boldsymbol{y}(k+i)$ and $\Delta\boldsymbol{u}(k+i)$ are the incremental form of predictive output and input vectors of the system in future time $k+i$ ($i$=1,2,⋯ ,$N$ ), respectively. Define $\boldsymbol{Y}_N(k)$, $\Delta\boldsymbol{Y}_N(k+1)$, $\Delta\boldsymbol{U}_N(k)$, $\Delta\boldsymbol{U}_{Nu}(k)$, $\boldsymbol{\Psi}_Y(k)$, $\tilde{\boldsymbol{\Psi}}_Y(k)$, $\boldsymbol{\Psi}_U(k)$, $\tilde{\boldsymbol{\Psi}}_U(k)$, $\boldsymbol{\Psi}_N(k)$ and $\tilde{\boldsymbol{\Psi}}_N(k)$ by the approximation $\boldsymbol{\phi}_{Ly}^T(k+i)=\boldsymbol{\phi}_{Ly}^T(k)$, $\boldsymbol{\phi}_{Lu}^T(k+i)=\boldsymbol{\phi}_{Lu}^T(k)$, ($i$=1,⋯, $N$-1) as follows, and define another set of $\boldsymbol{\Psi}_Y(k)$, $\tilde{\boldsymbol{\Psi}}_Y(k)$, $\boldsymbol{\Psi}_U(k)$, $\tilde{\boldsymbol{\Psi}}_U(k)$, $\boldsymbol{\Psi}_N(k)$ $\tilde{\boldsymbol{\Psi}}_N(k)$ in Appendix II if $\boldsymbol{\phi}_{Ly}^T(k+i)$ and $\boldsymbol{\phi}_{Lu}^T(k+i)$ can be obtained.

$$\Delta\boldsymbol{Y}_N(k+1)=\boldsymbol{Y}_N(k+1)-\boldsymbol{Y}_N(k)$$

$$\boldsymbol{Y}_N(k+1)=\begin{bmatrix}\boldsymbol{y}(k+1)\\ \vdots\\ \boldsymbol{y}(k+N)\end{bmatrix}_{N\bullet My\times 1}\qquad \boldsymbol{\Lambda}_N=\begin{bmatrix}\boldsymbol{I} & & \\ \vdots & \ddots & \\ \boldsymbol{I} & \cdots & \boldsymbol{I}\end{bmatrix}_{N\bullet My\times N\bullet My}$$

---

$$\boldsymbol{\Psi}_Y(k)=\begin{bmatrix}\boldsymbol{\varphi}_{11}\\ \boldsymbol{\varphi}_{12}\\ \boldsymbol{\varphi}_{13}\\ \boldsymbol{\varphi}_{14}\\ \vdots\\ \boldsymbol{\varphi}_{1N}\end{bmatrix}=\begin{bmatrix}\boldsymbol{\phi}_{Ly}^T(k)\\ \boldsymbol{\phi}_{Ly}^T(k)[\boldsymbol{C}+\boldsymbol{D}\boldsymbol{\phi}_{Ly}^T(k)]\\ \boldsymbol{\phi}_{Ly}^T(k)\boldsymbol{C}^2+\boldsymbol{\phi}_{Ly}^T(k)\sum_{i=0}^{1}\boldsymbol{C}^i\boldsymbol{D}\boldsymbol{\varphi}_{12-i}\\ \vdots\\ \boldsymbol{\phi}_{Ly}^T(k)\boldsymbol{C}^{N-1}\\ +\boldsymbol{\phi}_{Ly}^T(k)\sum_{i=0}^{N-2}\boldsymbol{C}^i\boldsymbol{D}\boldsymbol{\varphi}_{1N-i-1}\end{bmatrix}_{(N\bullet My)\times(Ly\bullet My)},$$

$$\tilde{\boldsymbol{\Psi}}_Y(k)=\boldsymbol{\Lambda}_N\tilde{\boldsymbol{\Psi}}_Y(k)=\begin{bmatrix}\boldsymbol{\phi}_{Ly}^T(k)\\ \boldsymbol{\phi}_{Ly}^T(k)[\boldsymbol{C}+\boldsymbol{D}\boldsymbol{\phi}_{Ly}^T(k)]\ +\boldsymbol{\phi}_{Ly}^T(k)\\ \vdots\\ \sum_{j=1}^{N}[\boldsymbol{\phi}_{Ly}^T(k)\boldsymbol{C}^{j-1}\\ +\boldsymbol{\phi}_{Ly}^T(k)\sum_{i=0}^{j-2}\boldsymbol{C}^i\boldsymbol{D}\boldsymbol{\varphi}_{1j-i-1}]\end{bmatrix}_{(N\bullet My)\times(Ly\bullet My)},$$

$$\begin{aligned}\boldsymbol{\Psi}_U(k)&=\begin{bmatrix}\boldsymbol{\varphi}_{21}^T & \boldsymbol{\varphi}_{22}^T & \boldsymbol{\varphi}_{23}^T & \cdots & \boldsymbol{\varphi}_{2N}^T\end{bmatrix}^T\\ &=\begin{bmatrix}\boldsymbol{\phi}_{Lu}^T(k)\boldsymbol{A}\\ \boldsymbol{\phi}_{Lu}^T(k)\boldsymbol{A}^2+\boldsymbol{\phi}_{Ly}^T(k)\boldsymbol{D}\boldsymbol{\phi}_{Lu}^T(k)\boldsymbol{A}\\ \boldsymbol{\phi}_{Lu}^T(k)\boldsymbol{A}^3+\boldsymbol{\phi}_{Ly}^T(k)\sum_{i=0}^{1}\boldsymbol{C}^i\boldsymbol{D}\boldsymbol{\varphi}_{22-i}\\ \vdots\\ \boldsymbol{\phi}_{Lu}^T(k)\boldsymbol{A}^N+\boldsymbol{\phi}_{Ly}^T(k)\sum_{i=0}^{N-2}\boldsymbol{C}^i\boldsymbol{D}\boldsymbol{\varphi}_{2N-i-1}\end{bmatrix},\\ &=[\boldsymbol{\Psi}_{U1}(k),\boldsymbol{\Psi}_{U2}(k),\cdots,\boldsymbol{\Psi}_{ULu-1}(k),\boldsymbol{0}]_{(N\bullet My)\times(Lu\bullet Mu)}\end{aligned}$$

$$\begin{aligned}\tilde{\boldsymbol{\Psi}}_U(k)&=\boldsymbol{\Lambda}_N\boldsymbol{\Psi}_U(k)\\ &=\begin{bmatrix}\boldsymbol{\phi}_{Lu}^T(k)\boldsymbol{A}\\ \boldsymbol{\phi}_{Lu}^T(k)\boldsymbol{A}^2+\boldsymbol{\phi}_{Ly}^T(k)\boldsymbol{D}\boldsymbol{\phi}_{Lu}^T(k)\boldsymbol{A}+\boldsymbol{\phi}_{Lu}^T(k)\boldsymbol{A}\\ \sum_{j=1}^{3}[\boldsymbol{\phi}_{Lu}^T(k)\boldsymbol{A}^j+\boldsymbol{\phi}_{Ly}^T(k)\sum_{i=0}^{j-2}\boldsymbol{C}^i\boldsymbol{D}\varphi_{2j-i-1}]\\ \vdots\\ \sum_{j=1}^{N}[\boldsymbol{\phi}_{Lu}^T(k)\boldsymbol{A}^j+\boldsymbol{\phi}_{Ly}^T(k)\sum_{i=0}^{j-2}\boldsymbol{C}^i\boldsymbol{D}\varphi_{2j-i-1}]\end{bmatrix},\\ &=[\tilde{\boldsymbol{\Psi}}_{U1}(k),\tilde{\boldsymbol{\Psi}}_{U2}(k),\cdots,\tilde{\boldsymbol{\Psi}}_{ULu-1}(k),\boldsymbol{0}]_{(N\bullet My)\times(Lu\bullet Mu)}\end{aligned}$$

$$\boldsymbol{\Psi}_N(k)_{(N\bullet My)\times(N\bullet Mu)}=\begin{bmatrix}\boldsymbol{\psi}_{11},\boldsymbol{\psi}_{12},\cdots,\boldsymbol{\psi}_{1N}\\ \boldsymbol{\psi}_{21},\boldsymbol{\psi}_{22},\cdots,\boldsymbol{\psi}_{2N}\\ \vdots\\ \boldsymbol{\psi}_{N2},\boldsymbol{\psi}_{N2},\cdots,\boldsymbol{\psi}_{NN}\end{bmatrix},\ \Delta\boldsymbol{U}_N(k)=\begin{bmatrix}\Delta\boldsymbol{u}(k)\\ \vdots\\ \Delta\boldsymbol{u}(k+N-1)\end{bmatrix}_{(N\bullet Mu)\times 1},\ \Delta\boldsymbol{U}_{Nu}(k)=\begin{bmatrix}\Delta\boldsymbol{u}(k)\\ \vdots\\ \Delta\boldsymbol{u}(k+N_u-1)\end{bmatrix}_{Nu\bullet Mu\times 1},$$

$$=\begin{bmatrix}\boldsymbol{\phi}_{Lu}^T(k)\boldsymbol{B} & \boldsymbol{0} & \boldsymbol{0} & \cdots & \boldsymbol{0}\\ \boldsymbol{\phi}_{Lu}^T(k)\boldsymbol{AB}+\boldsymbol{\phi}_{Ly}^T(k)\boldsymbol{D}\boldsymbol{\phi}_{Lu}^T(k)\boldsymbol{B} & \boldsymbol{\phi}_{Lu}^T(k)\boldsymbol{B} & \boldsymbol{0} & \cdots & \boldsymbol{0}\\ \begin{matrix}\boldsymbol{\phi}_{Lu}^T(k)\boldsymbol{A}^2\boldsymbol{B}\\ +\boldsymbol{\phi}_{Ly}^T(k)\sum_{i=0}^{1}\boldsymbol{C}^i\boldsymbol{D}\boldsymbol{\psi}_{2-i,1}\end{matrix} & \begin{matrix}\boldsymbol{\phi}_{Lu}^T(k)\boldsymbol{AB}\\ +\boldsymbol{\phi}_{Ly}^T(k)\boldsymbol{D}\boldsymbol{\phi}_{Lu}^T(k)\boldsymbol{B}\end{matrix} & \boldsymbol{\phi}_{Lu}^T(k)\boldsymbol{B} & \cdots & \boldsymbol{0}\\ \vdots & \vdots & \vdots & \ddots & \vdots\\ \begin{matrix}\boldsymbol{\phi}_{Lu}^T(k)\boldsymbol{A}^{N-1}\boldsymbol{B}\\ +\boldsymbol{\phi}_{Ly}^T(k)\sum_{i=0}^{N-2}\boldsymbol{C}^i\boldsymbol{D}\boldsymbol{\psi}_{N-i-1,1}\end{matrix} & \begin{matrix}\boldsymbol{\phi}_{Lu}^T(k)\boldsymbol{A}^{N-2}\boldsymbol{B}\\ +\boldsymbol{\phi}_{Ly}^T(k)\sum_{i=0}^{N-3}\boldsymbol{C}^i\boldsymbol{D}\boldsymbol{\psi}_{N-i-1,2}\end{matrix} & \begin{matrix}\boldsymbol{\phi}_{Lu}^T(k)\boldsymbol{A}^{N-3}\boldsymbol{B}\\ +\boldsymbol{\phi}_{Ly}^T(k)\sum_{i=0}^{N-4}\boldsymbol{C}^i\boldsymbol{D}\boldsymbol{\psi}_{N-i-1,3}\end{matrix} & \cdots & \boldsymbol{\phi}_{Lu}^T(k)\boldsymbol{B}\end{bmatrix},$$

$$\tilde{\boldsymbol{\Psi}}_N(k)_{(N\bullet My)\times(N\bullet Mu)}=\boldsymbol{\Lambda}_N\boldsymbol{\Psi}_N(k)$$

$$=\begin{bmatrix}
\boldsymbol{\phi}_{Lu}^T(k)\boldsymbol{B} & \boldsymbol{0} & \boldsymbol{0} & \cdots & \boldsymbol{0}\\
\begin{matrix}\boldsymbol{\phi}_{Lu}^T(k)\boldsymbol{AB}+\boldsymbol{\phi}_{Lu}^T(k)\boldsymbol{B}\\+\boldsymbol{\phi}_{Ly}^T(k)\boldsymbol{D}\boldsymbol{\phi}_{Lu}^T(k)\boldsymbol{B}\end{matrix} & \boldsymbol{\phi}_{Lu}^T(k)\boldsymbol{B} & \boldsymbol{0} & \cdots & \boldsymbol{0}\\
\begin{matrix}\sum_{j=1}^{3}[\boldsymbol{\phi}_{Lu}^T(k)\boldsymbol{A}^{j-1}\boldsymbol{B}\\+\boldsymbol{\phi}_{Ly}^T(k)\sum_{i=0}^{j-2}\boldsymbol{C}^i\boldsymbol{D}\boldsymbol{\psi}_{j-i-1,1}]\end{matrix} & \begin{matrix}\boldsymbol{\phi}_{Lu}^T(k)\boldsymbol{AB}+\boldsymbol{\phi}_{Lu}^T(k)\boldsymbol{B}\\+\boldsymbol{\phi}_{Ly}^T(k)\boldsymbol{D}\boldsymbol{\phi}_{Lu}^T(k)\boldsymbol{B}\end{matrix} & \boldsymbol{\phi}_{Lu}^T(k)\boldsymbol{B} & \cdots & \boldsymbol{0}\\
\vdots & \vdots & \vdots & \ddots & \vdots\\
\begin{matrix}\sum_{j=1}^{N}[\boldsymbol{\phi}_{Lu}^T(k)\boldsymbol{A}^{j-1}\boldsymbol{B}\\+\boldsymbol{\phi}_{Ly}^T(k)\sum_{i=0}^{j-2}\boldsymbol{C}^i\boldsymbol{D}\boldsymbol{\psi}_{j-i-1,1}]\end{matrix} & \begin{matrix}\sum_{j=2}^{N}[\boldsymbol{\phi}_{Lu}^T(k)\boldsymbol{A}^{j-2}\boldsymbol{B}\\+\boldsymbol{\phi}_{Ly}^T(k)\sum_{i=0}^{j-3}\boldsymbol{C}^i\boldsymbol{D}\boldsymbol{\psi}_{j-i-1.2}]\end{matrix} & \begin{matrix}\sum_{j=3}^{N}[\boldsymbol{\phi}_{Lu}^T(k)\boldsymbol{A}^{j-3}\boldsymbol{B}\\+\boldsymbol{\phi}_{Ly}^T(k)\sum_{i=0}^{j-4}\boldsymbol{C}^i\boldsymbol{D}\boldsymbol{\psi}_{j-i-1,3}]\end{matrix} & \cdots & \boldsymbol{\phi}_{Lu}^T(k)\boldsymbol{B}
\end{bmatrix},\ \boldsymbol{E}=\begin{bmatrix}\boldsymbol{I}\\\vdots\\\boldsymbol{I}\end{bmatrix}_{N\bullet My\times My},$$

$$\tilde{\boldsymbol{\Psi}}_{Nu}(k)_{\substack{(N\bullet My)\\\times(Nu\bullet Mu)}}=\begin{bmatrix}
\boldsymbol{\phi}_{Lu}^T(k)\boldsymbol{B} & \boldsymbol{0} & \cdots & \boldsymbol{0}\\
\boldsymbol{\phi}_{Lu}^T(k)\boldsymbol{AB}+\boldsymbol{\phi}_{Ly}^T(k)\boldsymbol{D}\boldsymbol{\phi}_{Lu}^T(k)\boldsymbol{B}+\boldsymbol{\phi}_{Lu}^T(k)\boldsymbol{B} & \boldsymbol{\phi}_{Lu}^T(k)\boldsymbol{B} & \cdots & \boldsymbol{0}\\
\vdots & \vdots & \ddots & \vdots\\
\sum_{j=1}^{Nu}[\boldsymbol{\phi}_{Lu}^T(k)\boldsymbol{A}^{j-1}\boldsymbol{B}+\boldsymbol{\phi}_{Ly}^T(k)\sum_{i=0}^{j-2}\boldsymbol{C}^i\boldsymbol{D}\boldsymbol{\psi}_{j-i-1,1}] & \begin{matrix}\sum_{j=2}^{Nu}[\boldsymbol{\phi}_{Lu}^T(k)\boldsymbol{A}^{j-2}\boldsymbol{B}\\+\boldsymbol{\phi}_{Ly}^T(k)\sum_{i=0}^{j-3}\boldsymbol{C}^i\boldsymbol{D}\boldsymbol{\psi}_{j-i-1,2}]\end{matrix} & \cdots & \boldsymbol{\phi}_{Lu}^T(k)\boldsymbol{B}\\
\vdots & \vdots & \vdots & \vdots\\
\sum_{j=1}^{N}[\boldsymbol{\phi}_{Lu}^T(k)\boldsymbol{A}^{j-1}\boldsymbol{B}+\boldsymbol{\phi}_{Ly}^T(k)\sum_{i=0}^{j-2}\boldsymbol{C}^i\boldsymbol{D}\boldsymbol{\psi}_{j-i-1,1}] & \begin{matrix}\sum_{j=2}^{N}[\boldsymbol{\phi}_{Lu}^T(k)\boldsymbol{A}^{j-2}\boldsymbol{B}\\+\boldsymbol{\phi}_{Ly}^T(k)\sum_{i=0}^{j-3}\boldsymbol{C}^i\boldsymbol{D}\boldsymbol{\psi}_{j-i-1,2}]\end{matrix} & \cdots & \begin{matrix}\sum_{j=Nu}^{N}[\boldsymbol{\phi}_{Lu}^T(k)\boldsymbol{A}^{j-Nu}\boldsymbol{B}\\+\boldsymbol{\phi}_{Ly}^T(k)\sum_{i=0}^{j-Nu-1}\boldsymbol{C}^i\boldsymbol{D}\boldsymbol{\psi}_{j-i-1,Nu}]\end{matrix}
\end{bmatrix}.$$

---

where $\boldsymbol{\varphi}_{1i}$ represents the rows from $(i-1)\bullet M_y+1$ to $i\bullet M_y$ in $\boldsymbol{\Psi}_Y(k)$; $\boldsymbol{\varphi}_{2i}$ represents the rows from $(i-1)\bullet M_y+1$ to $i\bullet M_y$ in $\boldsymbol{\Psi}_U(k)$, and $\boldsymbol{\Psi}_{Uj}(k)$ represents the columns from $(j-1)\bullet M_u+1$ to $j\bullet M_u$ in $\boldsymbol{\Psi}_U(k)$; $\tilde{\boldsymbol{\Psi}}_{Uj}(k)$ represents the columns from $(j-1)\bullet M_u+1$ to $j\bullet M_u$ in $\tilde{\boldsymbol{\Psi}}_{\mathrm{U}j}(k)$; $\boldsymbol{\psi}_{ij}$ represents the rows from $(i-1)\bullet M_y+1$ to $i\bullet M_y$ and columns from $(j-1)\bullet M_u+1$ to $j\bullet M_u$ in $\boldsymbol{\Psi}_N(k)$.

Then we can rewrite (4) as

$$\begin{aligned}\Delta\boldsymbol{Y}_N(k+1)&=\boldsymbol{\Psi}_Y(k)\Delta Y_{Ly}(k)+\boldsymbol{\Psi}_U(k)\Delta\boldsymbol{U}_{Lu}(k-1)\\&+\boldsymbol{\Psi}_N(k)\Delta\boldsymbol{U}_N(k)\end{aligned}\tag{5}$$

Both sides of (5) left multiply by $\boldsymbol{\Lambda}_N$ to obtain (6).

$$\begin{aligned}\boldsymbol{Y}_N(k+1)&=\boldsymbol{E}\boldsymbol{y}(k)+\boldsymbol{\Lambda}_N\boldsymbol{\Psi}_Y(k)\Delta\boldsymbol{Y}_{Ly}(k)+\boldsymbol{\Lambda}_N\boldsymbol{\Psi}_U(k)\Delta\boldsymbol{U}_{Lu}(k-1)\\&+\boldsymbol{\Lambda}_N\boldsymbol{\Psi}_N(k)\Delta\boldsymbol{U}_N(k)\\&=\boldsymbol{E}\boldsymbol{y}(k)+\tilde{\boldsymbol{\Psi}}_Y(k)\Delta\boldsymbol{Y}_{Ly}(k)+\tilde{\boldsymbol{\Psi}}_U(k)\Delta\boldsymbol{U}_{Lu}(k-1)\\&+\tilde{\boldsymbol{\Psi}}_N(k)\Delta\boldsymbol{U}_N(k)\end{aligned}\tag{6}$$

Define $N_u$ as control step length. Given $\Delta\boldsymbol{u}(k+j-1)=\boldsymbol{0}$, $N_u<j\le N$, (6) may be rewritten into

$$\begin{aligned}\boldsymbol{Y}_N(k+1)&=\boldsymbol{E}\boldsymbol{y}(k)+\tilde{\boldsymbol{\Psi}}_Y(k)\Delta\boldsymbol{Y}_{Ly}(k)+\tilde{\boldsymbol{\Psi}}_U(k)\Delta\boldsymbol{U}_{Lu}(k-1)\\&+\tilde{\boldsymbol{\Psi}}_{Nu}(k)\Delta\boldsymbol{U}_{Nu}(k)\end{aligned}\tag{7}$$

where $\tilde{\boldsymbol{\Psi}}_{Nu}(k)$ is defined as above.

## III. Model-Free Adaptive Predictive Control Design and Performance Analysis

### A. Design of MFAPC

We choose the following index function:

$$\begin{aligned}J=E\{&\left[\boldsymbol{Y}_N^*(k+1)-\boldsymbol{Y}_N(k+1)\right]^T\left[\boldsymbol{Y}_N^*(k+1)-\boldsymbol{Y}_N(k+1)\right]\\&+\Delta\boldsymbol{U}_{N_u}^T(k)\boldsymbol{\lambda}\Delta\boldsymbol{U}_{N_u}(k)\}\end{aligned}\tag{8}$$

where $\boldsymbol{\lambda}=dig(\lambda_1,\cdots,\lambda_{Mu\times N})$ is the weighted diagonal matrix with $\lambda_i$ ( $i=1,\cdots,M_u\times N$ ) equal to $\lambda$ according to [2]; $\tilde{\boldsymbol{Y}}_N^*(k+1)=\left[\boldsymbol{y}^*(k+1),\cdots,\boldsymbol{y}^*(k+N)\right]^T$ is the desired system output vector and $\boldsymbol{y}^*(k+i)=[y_1^*(k+i),\cdots,y_{My}^*(k+i)]$ is the desired output of system at time $k+i$ ($i$=1,2,⋯, $N$ ).

Inspired by the process of [31]-[33] and by substituting (7) into (8), we can obtain the controller by solving the optimal solution of (8) $\frac{\partial J}{\partial \Delta \boldsymbol{U}_{Nu}(k)}=\mathbf{0}$ and then organize it into

$$\begin{aligned}\Delta \boldsymbol{U}_{Nu}(k)=[&\tilde{\boldsymbol{\Psi}}_{Nu}^{T}(k)\tilde{\boldsymbol{\Psi}}_{Nu}(k)+\boldsymbol{\lambda}]^{-1}\tilde{\boldsymbol{\Psi}}_{Nu}^{T}(k)[(\boldsymbol{Y}_{N}^{*}(k+1)\\ &-\boldsymbol{E}\boldsymbol{y}(k))-\tilde{\boldsymbol{\Psi}}_{Y}(k)\Delta \boldsymbol{Y}_{Ly}(k)-\tilde{\boldsymbol{\Psi}}_{U}(k)\Delta \boldsymbol{U}_{Lu}(k-1)]\end{aligned} \tag{9}$$

Then the current input is given by

$$\boldsymbol{u}(k)=\boldsymbol{u}(k-1)+\boldsymbol{g}^{T}\Delta \boldsymbol{U}_{Nu}(k) \tag{10}$$

where $\boldsymbol{g}=\left[\boldsymbol{I},\mathbf{0},\cdots,\mathbf{0}\right]^{T}$.

## *B. Stability Analysis*

This section provides the performance analysis of MFAPC.

We define

$$\boldsymbol{\phi}_{Ly}(z^{-1})=\boldsymbol{\Phi}_{1}(k)+\cdots+\boldsymbol{\Phi}_{Ly}(k)z^{-Ly+1} \tag{11}$$

$$\boldsymbol{\phi}_{Lu}(z^{-1})=\boldsymbol{\Phi}_{Ly+1}(k)+\cdots+\boldsymbol{\Phi}_{Ly+Lu}(k)z^{-Lu+1} \tag{12}$$

$$\boldsymbol{P}(k)=[\tilde{\boldsymbol{\Psi}}_{Nu}^{T}(k)\tilde{\boldsymbol{\Psi}}_{Nu}(k)+\boldsymbol{\lambda}]^{-1}\tilde{\boldsymbol{\Psi}}_{Nu}^{T}(k) \tag{13}$$

, then (2) is rewritten as

$$\Delta \boldsymbol{y}(k+1)=\boldsymbol{\phi}_{Ly}(z^{-1})\Delta \boldsymbol{y}(k)+\boldsymbol{\phi}_{Lu}(z^{-1})\Delta \boldsymbol{u}(k) \tag{14}$$

where $z^{-1}$ is the backward shift operator and $\Delta=1-z^{-1}$.

We combine (9), (10) and (14) to obtain

$$\begin{aligned}&\Big[(\boldsymbol{I}-z^{-1}\boldsymbol{\phi}_{Ly}(z^{-1}))\Delta+z^{-1}\boldsymbol{\phi}_{Lu}(z^{-1})(\boldsymbol{I}+z^{-1}\boldsymbol{P}(k)\tilde{\boldsymbol{\Psi}}_{U}(k)\boldsymbol{T}_{u})^{-1}\boldsymbol{P}(k)\\ &\bullet\tilde{\boldsymbol{\Psi}}_{Y}(k)\boldsymbol{T}_{y}\Delta+z^{-1}\boldsymbol{\phi}_{Lu}(z^{-1})(\boldsymbol{I}+z^{-1}\boldsymbol{P}(k)\tilde{\boldsymbol{\Psi}}_{U}(k)\boldsymbol{T}_{u})^{-1}\boldsymbol{P}(k)\boldsymbol{E}\Big]\boldsymbol{y}(k+1)\\ &=\boldsymbol{\phi}_{Lu}(z^{-1})(\boldsymbol{I}+z^{-1}\boldsymbol{P}(k)\tilde{\boldsymbol{\Psi}}_{U}(k)\boldsymbol{T}_{u})^{-1}\boldsymbol{P}(k)\boldsymbol{H}\boldsymbol{y}^{*}(k+1)\end{aligned} \tag{15}$$

where $\boldsymbol{H}=[\boldsymbol{I},z\boldsymbol{I},\cdots,z^{N-1}\boldsymbol{I}]^{T}$ , $\boldsymbol{T}_{y}=[\boldsymbol{I},\cdots,z^{-Ly+1}\boldsymbol{I}]^{T}$ and $\boldsymbol{T}_{u}=[\boldsymbol{I},z^{-1}\boldsymbol{I},\cdots,z^{-Lu+1}\boldsymbol{I}]^{T}$.

If $\operatorname{rank}\left[\boldsymbol{\phi}_{Lu}(z^{-1})\right]=M_{y}$ ( $M_{u}\geq M_{y}$ ), we may choose appropriate $N$, $N_u$ and $\boldsymbol{\lambda}$ such that the following inequality:

$$\begin{aligned}\boldsymbol{T}(z^{-1})=&\Big[(\boldsymbol{I}-z^{-1}\boldsymbol{\phi}_{Ly}(z^{-1}))\Delta\\ &+z^{-1}\boldsymbol{\phi}_{Lu}(z^{-1})(\boldsymbol{I}+z^{-1}\boldsymbol{P}(k)\tilde{\boldsymbol{\Psi}}_{U}(k)\boldsymbol{T}_{u})^{-1}\boldsymbol{P}(k)\tilde{\boldsymbol{\Psi}}_{Y}(k)\boldsymbol{T}_{y}\Delta\\ &+z^{-1}\boldsymbol{\phi}_{Lu}(z^{-1})(\boldsymbol{I}+z^{-1}\boldsymbol{P}(k)\tilde{\boldsymbol{\Psi}}_{U}(k)\boldsymbol{T}_{u})^{-1}\boldsymbol{P}(k)\boldsymbol{E}\Big]\neq\mathbf{0}\quad |z|>1\end{aligned} \tag{16}$$

for the system stability, according to [28] and [33].

Further, if the system stability is guaranteed, sometimes we may determine the steady-state error vector by

$$\begin{aligned}&\lim_{k\to\infty}\boldsymbol{e}(k)\\ &=\lim_{z\to 1}(1-z^{-1})\boldsymbol{T}^{-1}(z^{-1})\Big[(\boldsymbol{I}-z^{-1}\boldsymbol{\phi}_{Ly}(z^{-1}))\Delta\\ &\quad+z^{-1}\boldsymbol{\phi}_{Lu}(z^{-1})(\boldsymbol{I}+z^{-1}\boldsymbol{P}(k)\tilde{\boldsymbol{\Psi}}_{U}(k)\boldsymbol{T}_{u})^{-1}\boldsymbol{P}(k)\tilde{\boldsymbol{\Psi}}_{Y}(k)\boldsymbol{T}_{y}\Delta\\ &\quad-\boldsymbol{\phi}_{Lu}(z^{-1})(\boldsymbol{I}+z^{-1}\boldsymbol{P}(k)\tilde{\boldsymbol{\Psi}}_{U}(k)\boldsymbol{T}_{u})^{-1}\boldsymbol{P}(k)(\boldsymbol{H}-z^{-1}\boldsymbol{E})\Big]\boldsymbol{Z}(\boldsymbol{y}^{*}(k+1))\end{aligned} \tag{17}$$

In applications, we normally design the MFAC controller by increasing $\boldsymbol{\lambda}$ for the robustness of systems and decreasing $\boldsymbol{\lambda}$ for the convergent performance of systems. When the desired trajectories are step signals $\boldsymbol{y}^{*}(k+i)=[1,\cdots,1]_{1\times My}^{T}$ ($i$=1, ···, $N$), we may eliminate the steady-state error by choosing $\boldsymbol{\lambda}=\mathbf{0}$ if the system stability is guaranteed.

## IV. SIMULATIONS

Example 1.1: In this example, we suppose the offline model is established precisely for the study and to make a comparison between MFAPC and MFAC. The model is given as the following linear system:

$$\begin{aligned}&\boldsymbol{y}(k+1)=\boldsymbol{\Phi}_{1}(k)\boldsymbol{y}(k)+\boldsymbol{\Phi}_{2}(k)\boldsymbol{u}(k)+\boldsymbol{\Phi}_{3}(k)\boldsymbol{u}(k-1)\\ &=\begin{bmatrix}-1 & 2\\ -1 & 1.4\end{bmatrix}\begin{bmatrix}y_{1}(k)\\ y_{2}(k)\end{bmatrix}+\begin{bmatrix}1.3 & 0\\ 1 & 0\end{bmatrix}\begin{bmatrix}u_{1}(k)\\ u_{2}(k)\end{bmatrix}+\begin{bmatrix}0.7 & 0.5\\ 0.6 & 0.8\end{bmatrix}\begin{bmatrix}u_{1}(k-1)\\ u_{2}(k-1)\end{bmatrix}\end{aligned} \tag{18}$$

The minimum time delay between $\boldsymbol{y}(k+1)$ and $\boldsymbol{u}(k)$ is $\boldsymbol{d}=\begin{bmatrix}d_{11} & d_{12}\\ d_{21} & d_{22}\end{bmatrix}=\begin{bmatrix}1 & 2\\ 1 & 2\end{bmatrix}$. The desired output trajectories are

$$y_{1}^{*}(k+1)=y_{2}^{*}(k+1)=3\times(-1)^{round(k/50)} \tag{19}$$

The initial values are $\boldsymbol{y}(1)=\boldsymbol{y}(3)=[0,0]^{T}$ , $\boldsymbol{y}(2)=[1,1]^{T}$ , which are cited from [2]. The controller structure is applied with $L_{y}=n_{y}+1=1$, $L_{u}=n_{u}+1=2$. And we choose $\boldsymbol{\lambda}_{MFAPC}=10^{-4}\boldsymbol{I}$, $\boldsymbol{\lambda}_{MFAC}=10^{-3}\boldsymbol{I}$, $N=2$ and $N_{u}=2$.

The outputs of the system controlled by the proposed MFAPC and current corrected MFAC are shown in Fig. 1 and Fig. 2, respectively. The outputs of the two controllers are shown in Fig. 3.

Based on the analysis of Fig. 1 and Fig. 2, it is evident that the system's tracking performance is excellent when controlled by the proposed MFAPC. Both controllers are essentially optimal solutions of the index function (8) and aim to achieve optimal performance in the control process. However, the static error is hard to eliminate when the system is controlled by the current MFAC, regardless of the chosen values for $\boldsymbol{\lambda}_{MFAC}$. The existence of two-time delays between $u_{2}(k)$ and $\boldsymbol{y}(k+1)$ results in the absence of $u_{2}(k)$ in the controller. [28] imposes the interactor matrix in almost every adaptive control method to solve this problem of the different delay times among the inputs and outputs of multivariable systems. Unfortunately, the controller structure in [28] needs to be adjusted manually based on the interactor matrix when the time delays change.

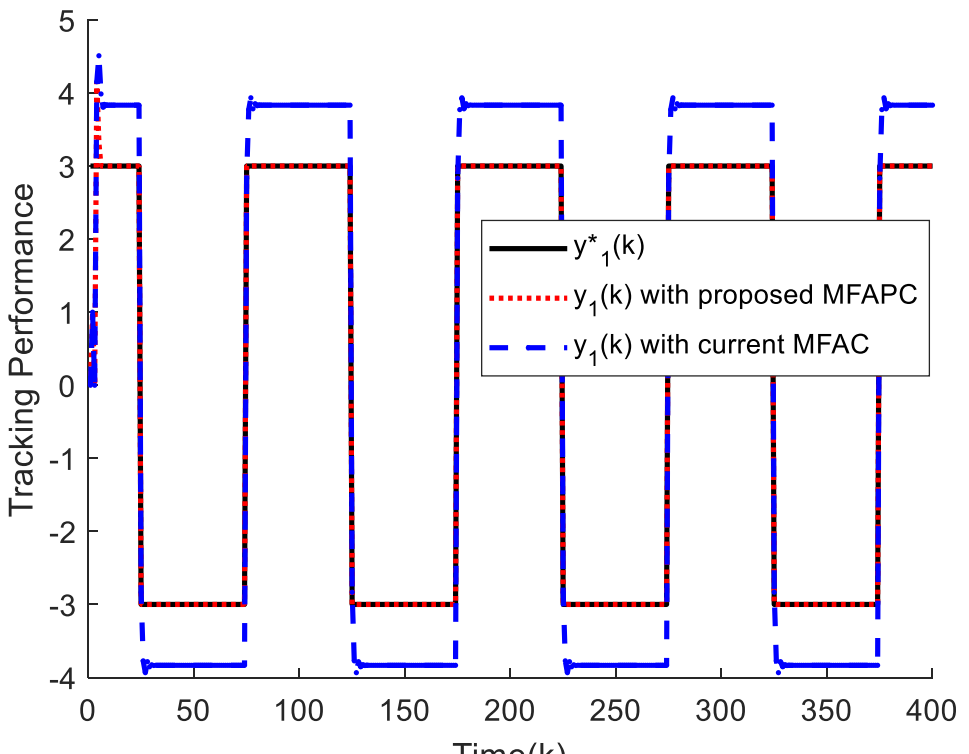

Fig. 1 Tracking performance of $y_1$

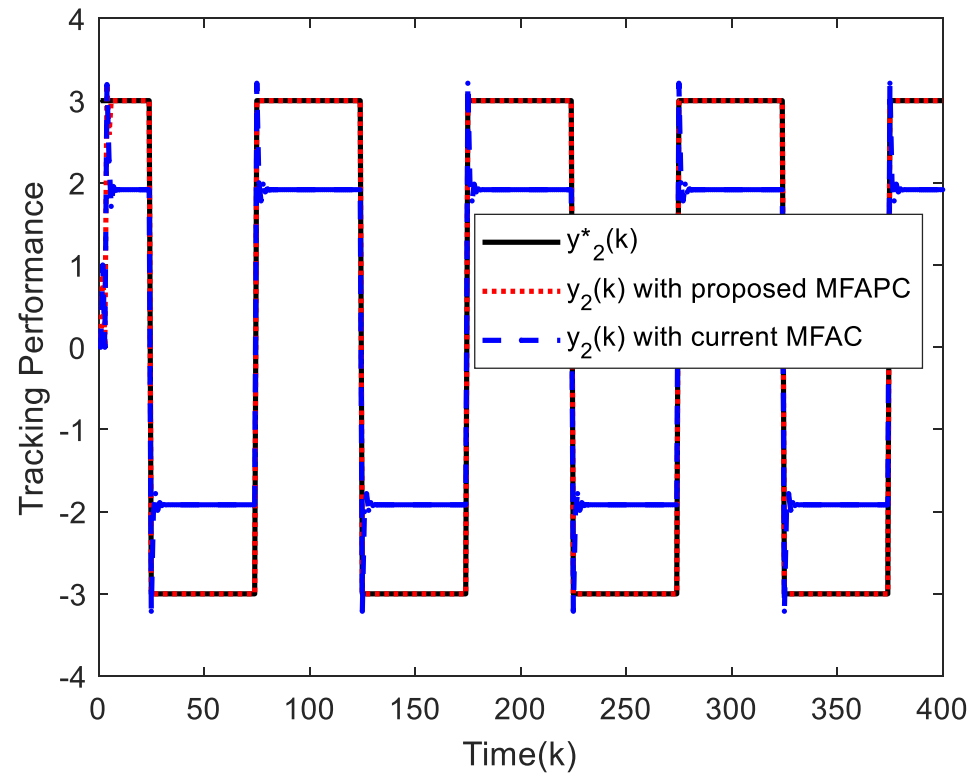

Fig. 2 Tracking performance of $y2$

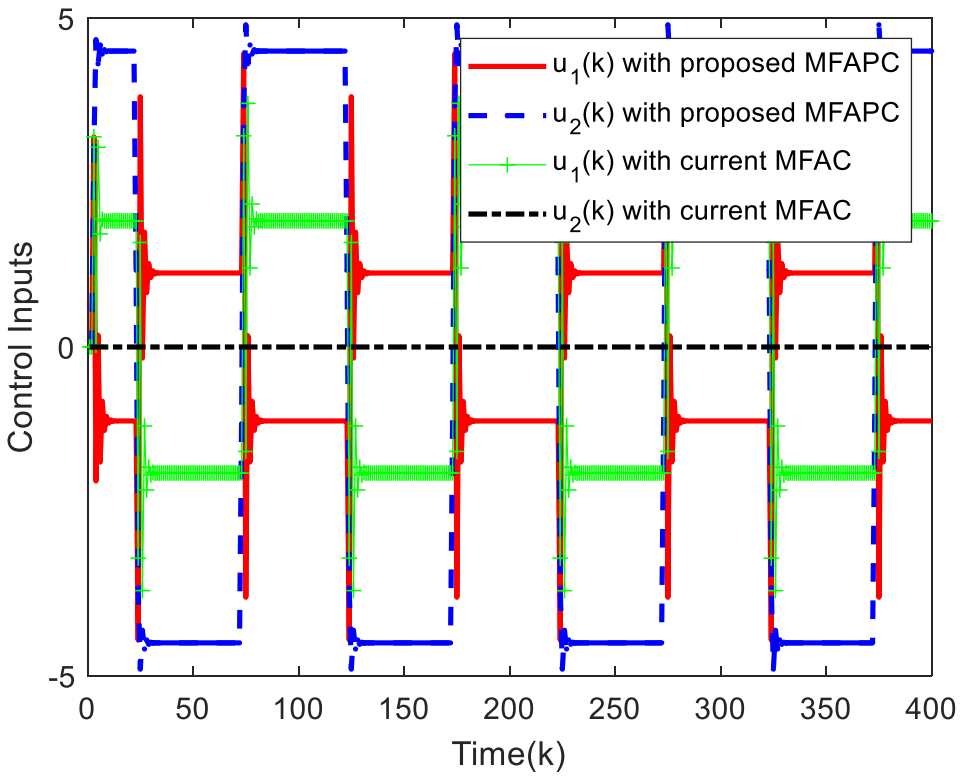

Fig. 3 Control inputs

Fig. 4 shows the tracking performance of $y_1(k)$ when there is constant disturbance $\boldsymbol{w}(k)=[5 \quad 10]^T$ in the system (18).

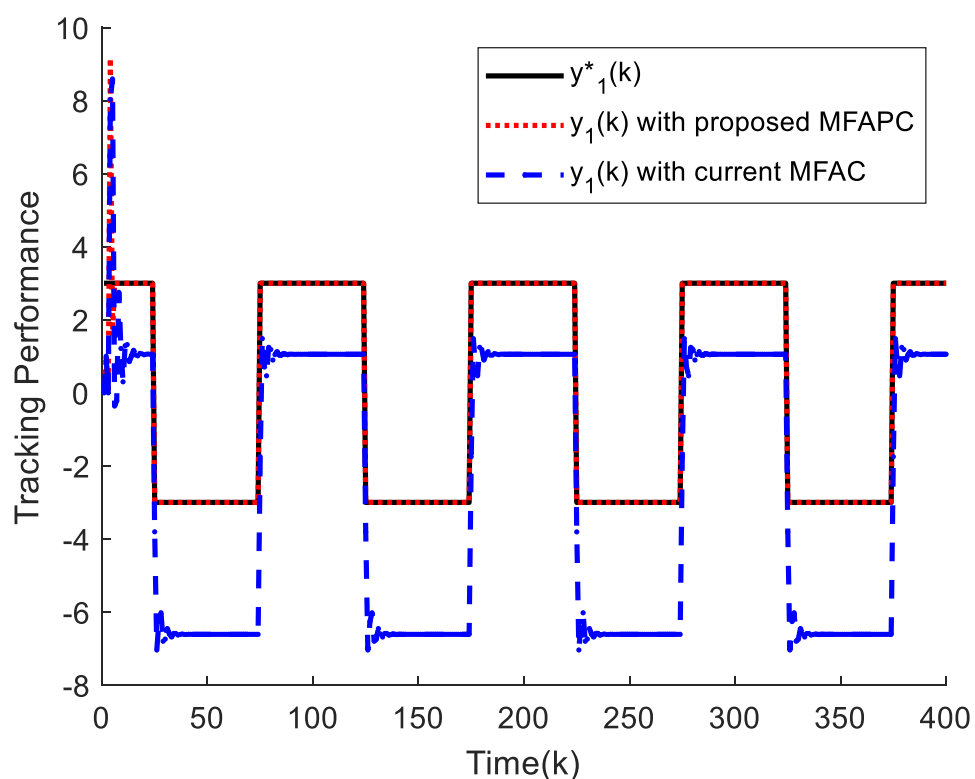

Fig. 4 Tracking performance of $y_1$

We can see that the static error is also eliminated by the inherent integrator in MFAPC [34].

Example 1.2: In this example, the model is given as the following linear system with $M_y$=2, $M_u$=3:

$$\boldsymbol{y}(k+1)=\boldsymbol{\Phi}_1(k)\boldsymbol{y}(k)+\boldsymbol{\Phi}_2(k)\boldsymbol{u}(k)+\boldsymbol{\Phi}_3(k)\boldsymbol{u}(k-1)$$

$$=\begin{bmatrix}-1 & 1\\ -1 & 1\end{bmatrix}\begin{bmatrix}y_1(k)\\ y_2(k)\end{bmatrix}+\begin{bmatrix}0 & 0 & 0\\ 0 & 0 & 0\end{bmatrix}\begin{bmatrix}u_1(k)\\ u_2(k)\\ u_3(k)\end{bmatrix} \tag{20}$$

$$+\begin{bmatrix}0.7 & 0.2 & 0.4\\ 0.6 & 0.8 & 0.4\end{bmatrix}\begin{bmatrix}u_1(k-1)\\ u_2(k-1)\\ u_3(k-1)\end{bmatrix}$$

The initial values are $\boldsymbol{y}(1)=\boldsymbol{y}(3)=[0,0]^T$, $\boldsymbol{y}(2)=[1,1]^T$, $\boldsymbol{u}(1)=\boldsymbol{u}(2)=[0,0,0]^T$. The controller parameters are $\boldsymbol{\lambda}_{MFAPC}=0.01\boldsymbol{I}$, $\boldsymbol{\lambda}_{MFAC}=\boldsymbol{I}$. All the other settings are the same as in Example 1.1.

Fig. 5 and Fig. 6 show the tracking performance of the system controlled by MFAPC and MFAC, respectively. Fig. 7 shows the outputs of MFAPC and MFAC. From Fig. 5 and Fig. 6, we can see that the outputs of the system controlled by MFAC are zero. The reason is that the existing time delay $\boldsymbol{d}=\begin{bmatrix}2 & 2 & 2\\ 2 & 2 & 2\end{bmatrix}$ in the system causes the absence of MFAC controller outputs.

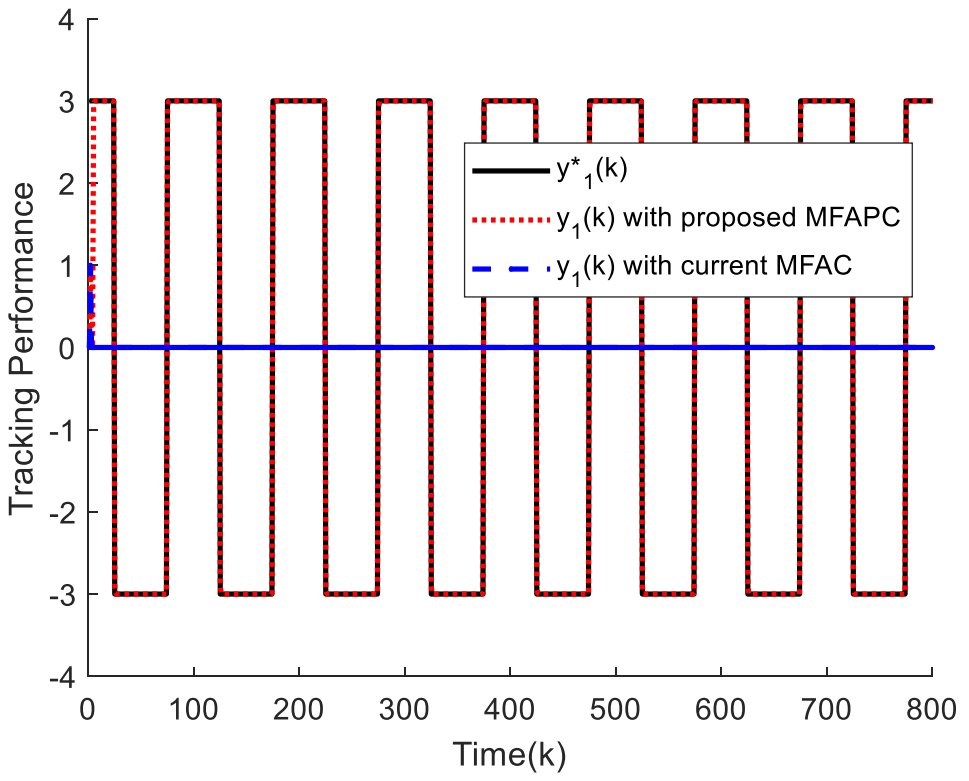

Fig. 5 Tracking performance of $y_1$

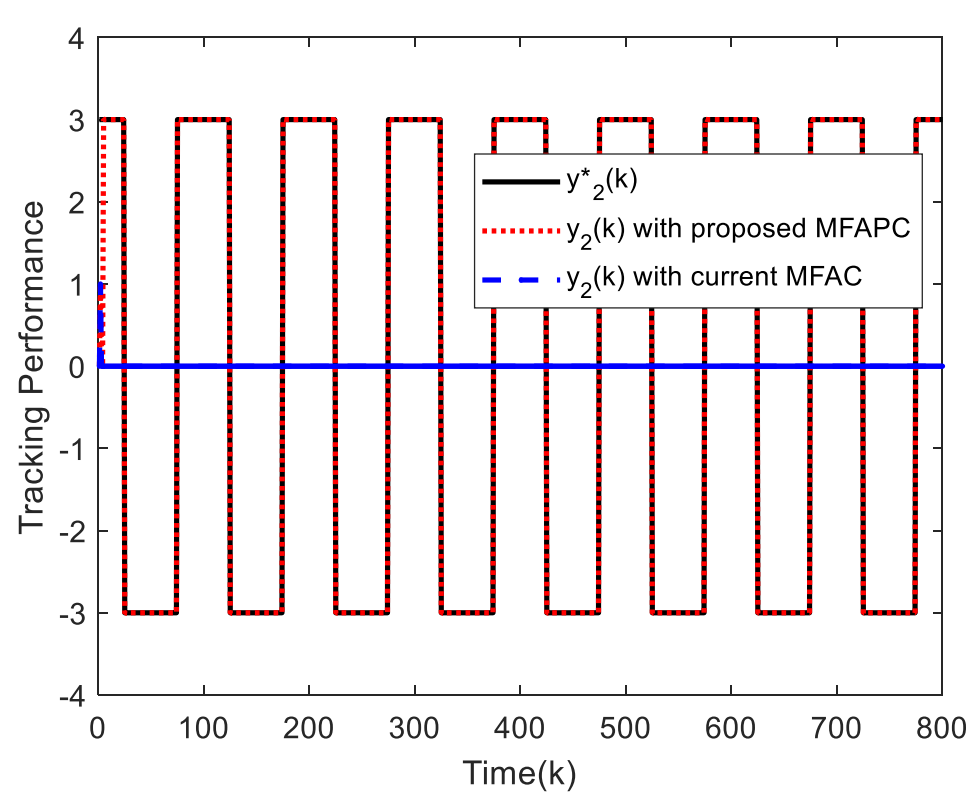

Fig. 6 Tracking performance of $y2$

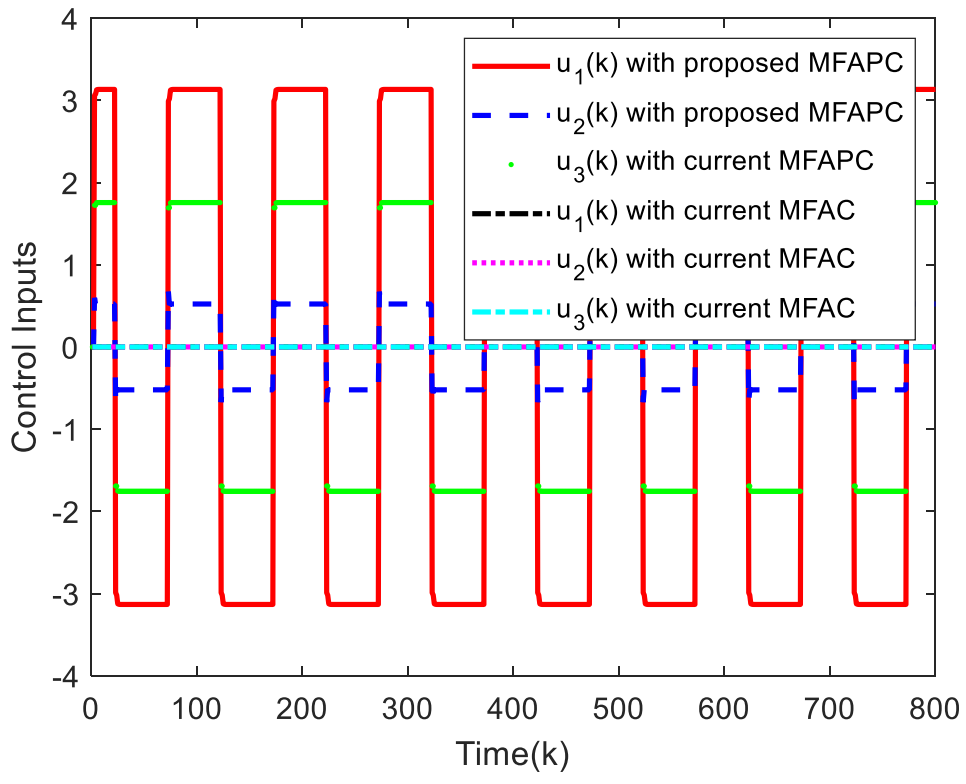

Fig. 7 Control inputs

Example 1.3: In this example, we suppose the time delay of the system in Example 1.2 is unknown. We introduce the online projection identification method [1], [2] for further studying MFAPC and MFAC.

All the settings are the same as in Example 1.2. The projection identification method in [2] is corrected into

$$\boldsymbol{\phi}_L^T(k) = \boldsymbol{\phi}_L^T(k-1) + \eta\left[\boldsymbol{y}(k) - \boldsymbol{y}(k-1) - \boldsymbol{\phi}_L^T(k-1)\Delta\boldsymbol{H}(k-1)\right]\bullet$$
$$\Delta\boldsymbol{H}^T(k-1)\left[\mu\boldsymbol{I} + \Delta\boldsymbol{H}(k-1)\Delta\boldsymbol{H}^T(k-1)\right]^{-1} \tag{21}$$

The parameters of the projection identification method are set to $\eta = 1.5$ and $\mu = 1$ [2]. The initial value of PJM is $\hat{\boldsymbol{\phi}}_L^T(1) = \hat{\boldsymbol{\phi}}_L^T(2) = \hat{\boldsymbol{\phi}}_L^T(3) = 0.01\bullet\begin{bmatrix} 1 & 1 & 1 & 1 & 1 & 1 & 1 & 1 & 1 \\ 1 & 1 & 1 & 1 & 1 & 1 & 1 & 1 & 1 \end{bmatrix}$ [35], which implies that the time delay is not figured out.

Fig. 8 and Fig. 9 respectively show the tracking performance of the system controlled by MFAPC and MFAC. Fig. 10 shows the outputs of MFAPC and MFAC. Fig. 11 and Fig. 12 show the elements of the estimated PJM corresponding to MFAPC and MFAC, respectively.

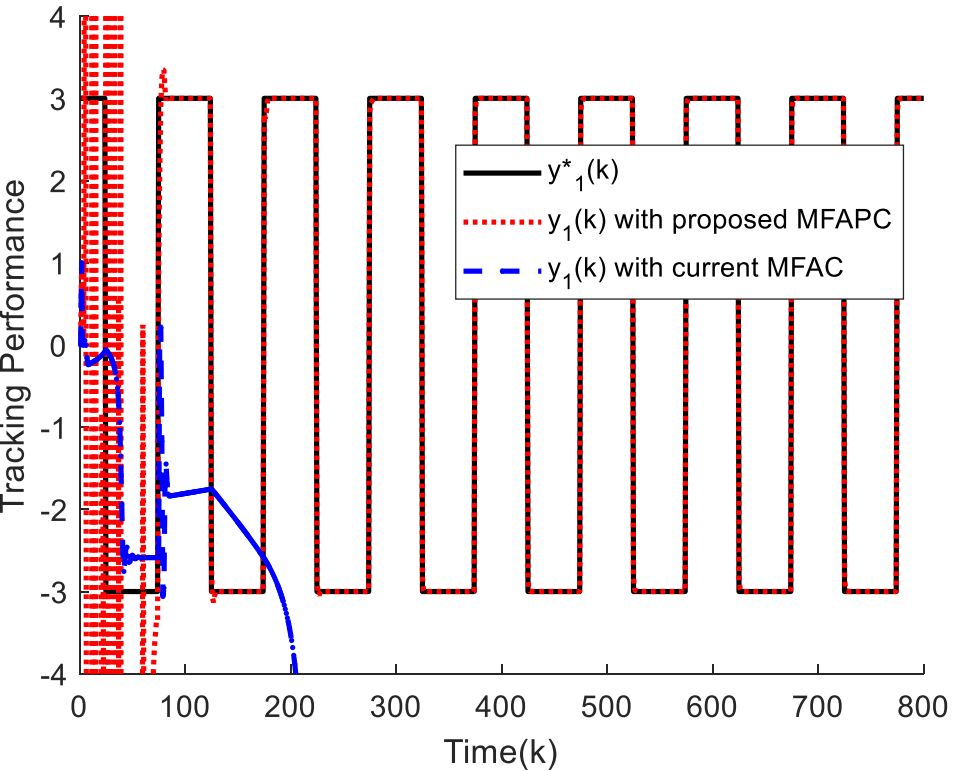

Fig. 8 Tracking performance of $y_1$

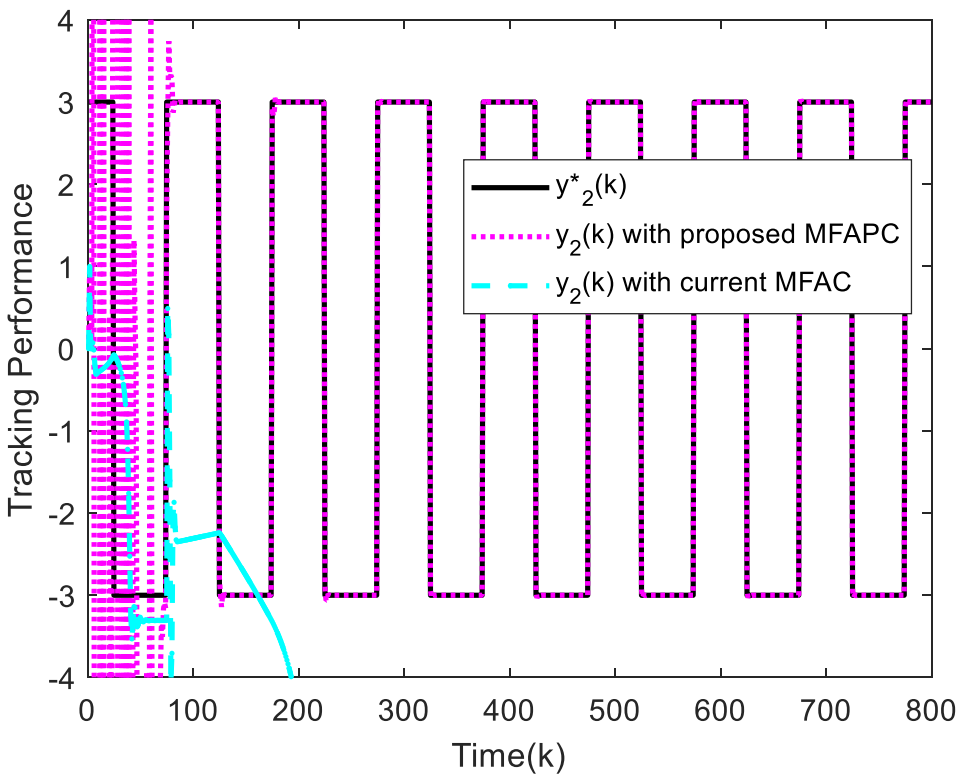

Fig. 9 Tracking performance of $y2$

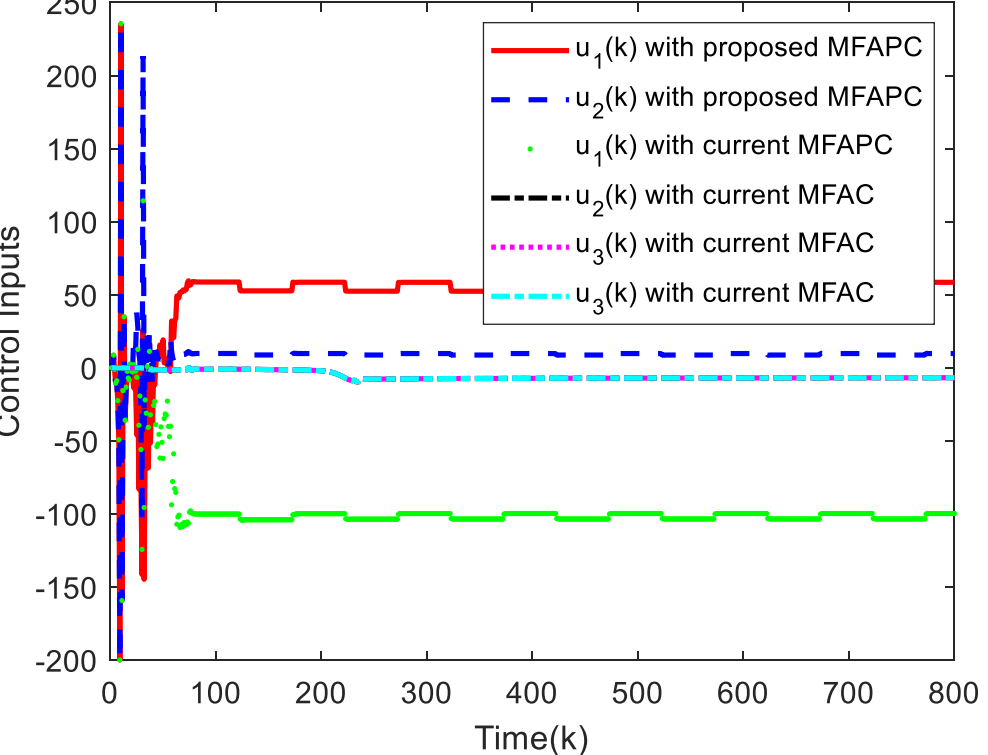

Fig. 10 Control inputs

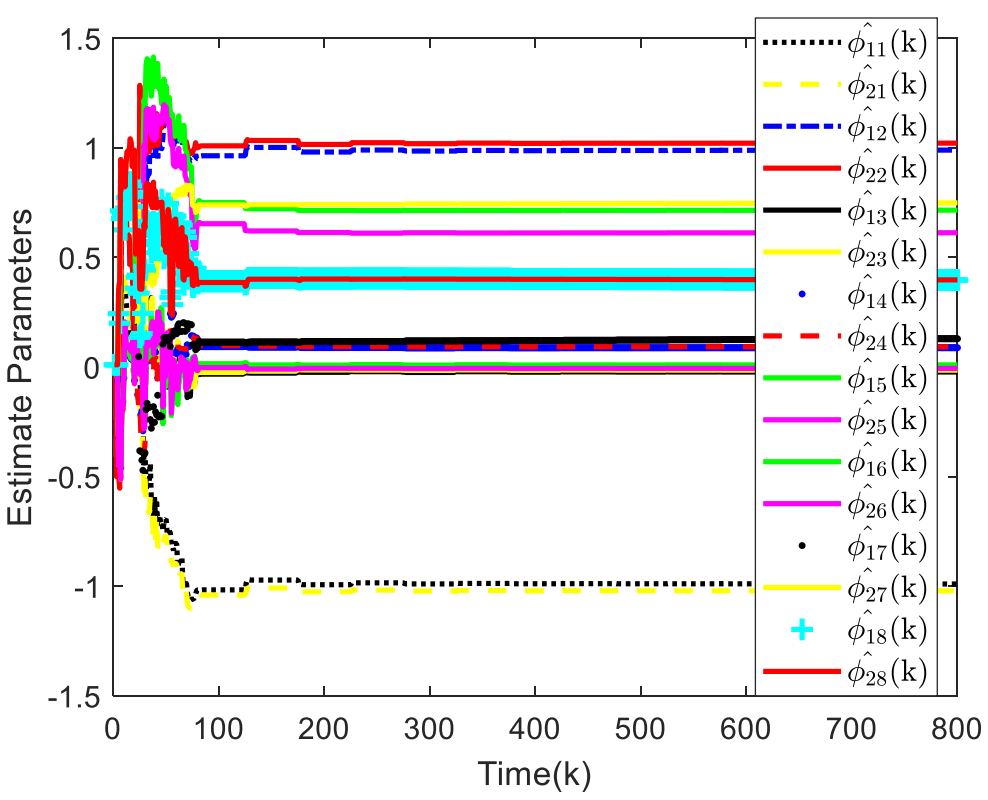

Fig. 11 Estimated PJM of MFAPC

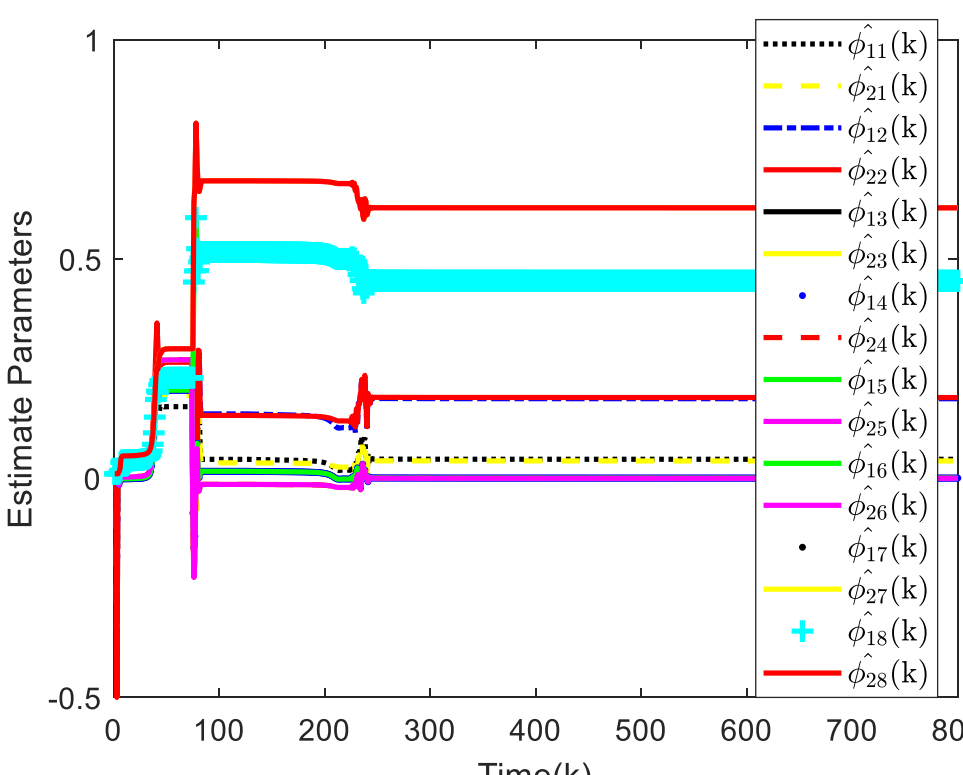

Fig. 12 Estimated PJM of MFAC

It is interesting to note that Fig. 11 and Fig. 12 present different results regarding the estimated elements in PJM when the system is controlled by MFAPC and MFAC separately, even though we used the same online identification method and parameters.

The system controlled by MFAC is performing poorly. Similarly, the tracking performance of the system controlled by MFAPC is also subpar in the time interval of [1,100], since the estimated PJM fails to provide an accurate reflection of the system. After the time of 100, the estimated elements in PJM are relatively stable and meanwhile the system is stable with the convergence of tracking error. However, the tracking performance is inferior to that in Example 1.2 since the established model lacks precision. Nevertheless, it provides a relatively objective representation of the system through a set of estimated elements in PJM after the time of 100. Without loss of generality, assuming the experimental system model is (20), we normally fix the coefficients in MFAPC controller in accordance with the roughly online estimated PJM in the time of 800 to have an acceptable performance in practice.

Example 2: In this example, we want to show how to apply the MFAPC controller in nonlinear systems in the right way. The model is given as the following nonlinear system:

$$\begin{bmatrix} y_1(k+1) \\ y_2(k+1) \end{bmatrix} = \begin{bmatrix} -0.1y_1^3(k)+0.1y_2^2(k)+0.7u_1(k-1)+0.5u_2(k-1) \\ -0.1y_1^2(k)+0.2y_2^3(k)+0.6u_1(k-1)+0.8u_2(k-1) \end{bmatrix.}$$
$$\begin{matrix} +0.2u_1^3(k)+\cos u_1^2(k)+0.1u_2^3(k)+0.5\sin u_2^2(k) \\ +0.1u_1^4(k)+0.2\sin(u_1(k))+0.1u_2^2(k)+0.9u_2(k) \end{matrix}\Bigg] \tag{22}$$

The desired output trajectories are

$$\begin{aligned} & y_1^*(k)=5\sin(k/40)+2\cos(k/20) && 1\le k\le 400 \\ & y_2^*(k)=2\sin(k/10)+5\sin(k/30) && 1\le k\le 400 \\ & y_1^*(k)=y_2^*(k)=(-1)^{round(k/50)} && 401\le k\le 800 \end{aligned} \tag{23}$$

The initial values are $\boldsymbol{y}(1)=\boldsymbol{y}(3)=\boldsymbol{0}$, $\boldsymbol{y}(2)=[1,1]^T$, $\boldsymbol{u}(1)=\boldsymbol{u}(2)=[0,0]^T$, $\boldsymbol{\phi}_L^T(1)=\boldsymbol{\phi}_L^T(2)=0.01\bullet\begin{bmatrix} 1 & 1 & 1 & 1 & 1 & 1 \\ 1 & 1 & 1 & 1 & 1 & 1 \end{bmatrix}$.

The controller structure is applied with $L_y=n_y+1=1$, $L_u=n_u+1=2$. And we choose $\boldsymbol{\lambda}_{MFAPC}=\boldsymbol{I}$, $\boldsymbol{\lambda}_{MFAC}=33\boldsymbol{I}$, $N$=2 and $N_u$=2. The elements in PJM are calculated by

$$\boldsymbol{\phi}_{Ly}^T(k)=\boldsymbol{\Phi}_1(k)=\frac{\partial \boldsymbol{f}(\boldsymbol{\varphi}(k-1))}{\partial \boldsymbol{y}^T(k-1)}=\begin{bmatrix} -0.3y_1^2(k-1) & 0.2y_2(k-1) \\ -0.2y_1(k-1) & 0.6y_2^2(k-1) \end{bmatrix},$$

$$\boldsymbol{\Phi}_2(k)=\frac{\partial \boldsymbol{f}(\boldsymbol{\varphi}(k-1))}{\partial \boldsymbol{u}^T(k-1)}=\begin{bmatrix} 0.6u_1^2(k-1) & 0.3u_2^2(k-1) \\ 0.4u_1^3(k-1) & 0.2u_2(k-1) \end{bmatrix} + \begin{bmatrix} -2u_1(k-1)\sin(u_1^2(k-1)) & u_2(k-1)\cos(u_2^2(k-1)) \\ 0.2\cos(u_1(k-1)) & 0.9 \end{bmatrix},$$

$$\boldsymbol{\Phi}_3(k)=\frac{\partial \boldsymbol{f}(\boldsymbol{\varphi}(k-1))}{\partial \boldsymbol{u}^T(k-2)}=\begin{bmatrix} 0.7 & 0.5 \\ 0.6 & 0.8 \end{bmatrix},\quad \boldsymbol{\phi}_{Lu}^T(k)=[\boldsymbol{\Phi}_2(k),\boldsymbol{\Phi}_3(k)]$$

The outputs of the system controlled by the proposed MFAPC and current corrected MFAC in [23] are shown in Fig. 13 and Fig. 14, respectively. The outputs of two controllers are shown in Fig. 15. Fig. 16 shows the calculated elements in PJM for MFAPC.

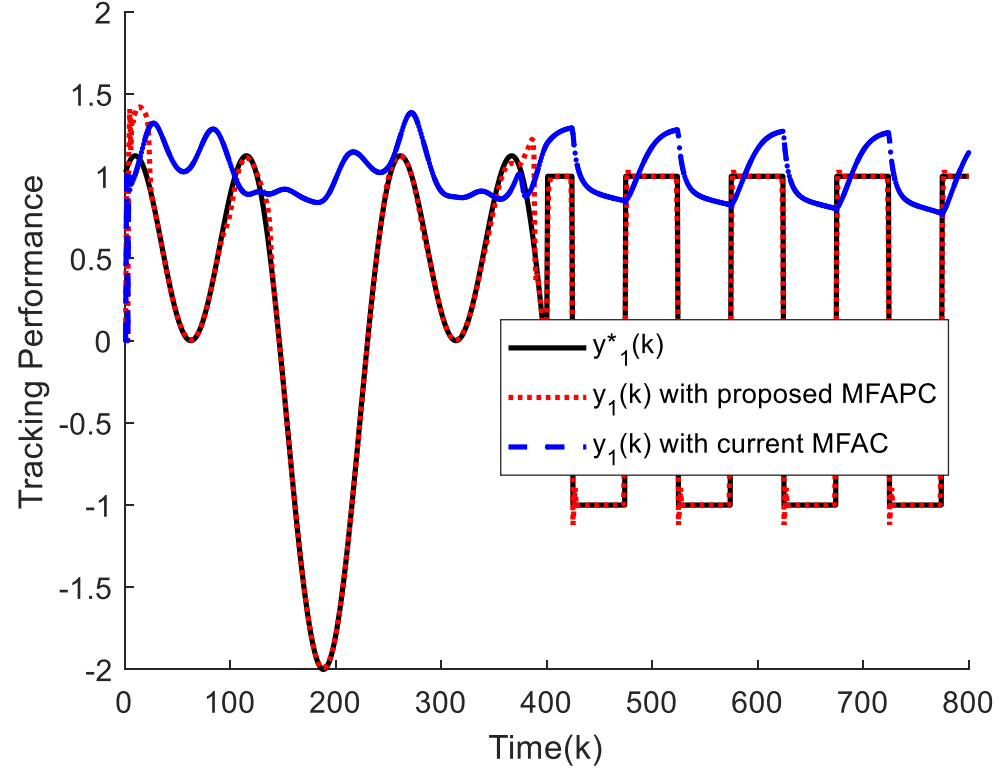

Fig. 13 Tracking performance of $y_1$

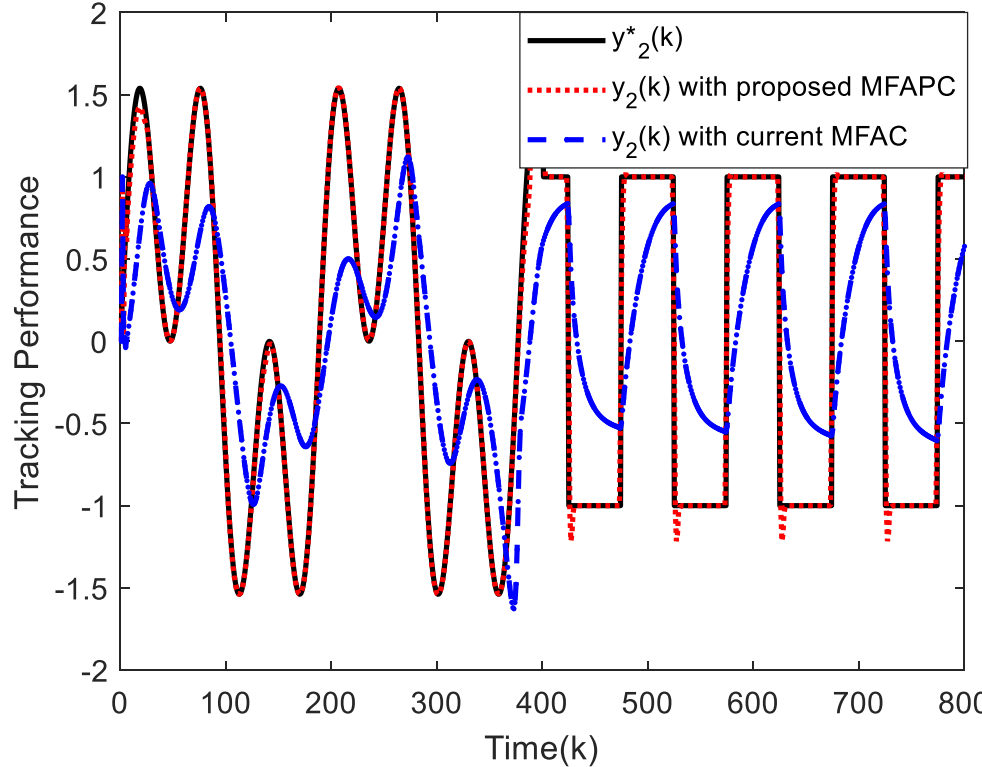

Fig. 14 Tracking performance of $y2$

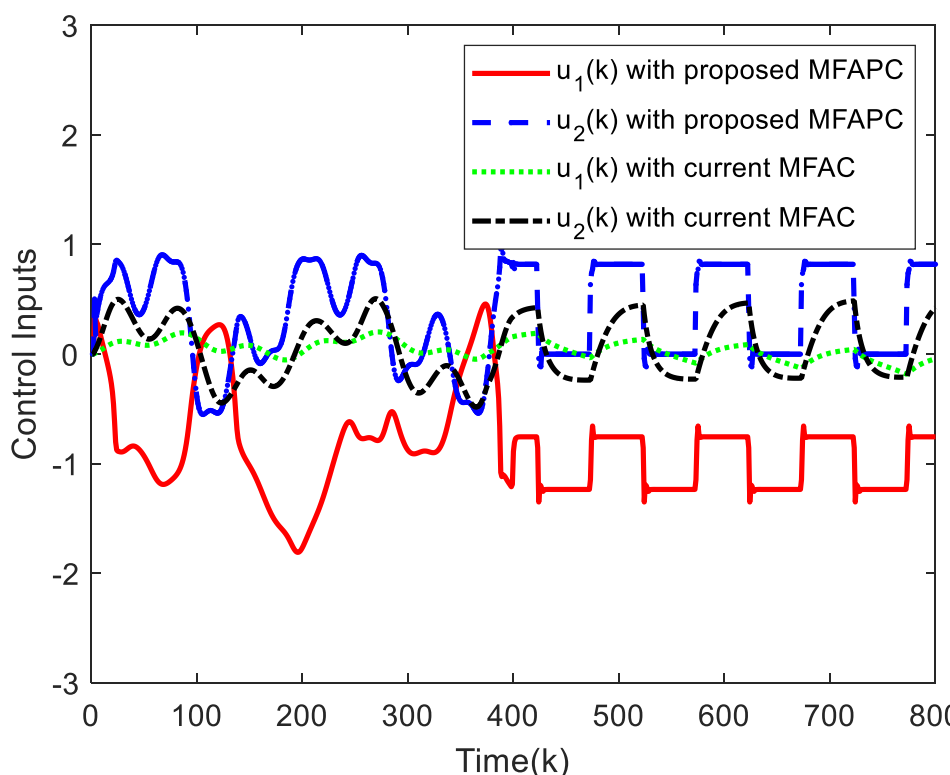

Fig. 15 Control inputs

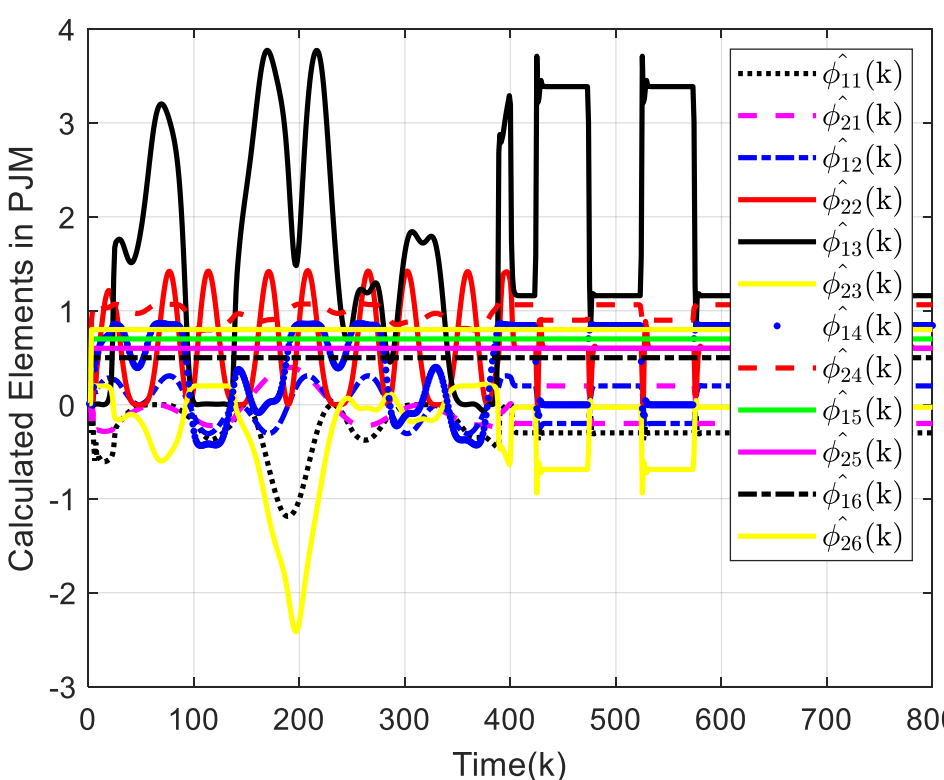

Fig. 16 Calculated elements in PJM of MFAPC

From Fig. 13 and Fig. 14, we know that the system controlled by MFAPC can track the trajectories. However, the behavior of system controlled by MFAC is unacceptable, and the system will be divergent if $\boldsymbol{\lambda}$ <32. This discrepancy might be attributed to that $[\tilde{\boldsymbol{\Psi}}_{Nu}^T(k)\tilde{\boldsymbol{\Psi}}_{Nu}(k)+\lambda\boldsymbol{I}]^{-1}\tilde{\boldsymbol{\Psi}}_{Nu}^T(k)$ in MFAPC is more stable than $[\boldsymbol{\Phi}_{Ly+1}^T(k)\boldsymbol{\Phi}_{Ly+1}(k)+\lambda\boldsymbol{I}]^{-1}\boldsymbol{\Phi}_{Ly+1}^T(k)$ in MFAC. Besides, this example also shows we'd better choose $L_y=n_y+1$ and $L_u=n_u+1$ in practice if $n_y$ and $n_u$ can be obtained. However, this significant conclusion contradicts the current works on MFAC whose fundamental tools, i.e., EDLM, are based on $1\le L_y\le n_y$, $1\le L_u\le n_u$. To this end, we extend the range to $0\le L_y$ and $1\le L_u$, as proven in Appendix I.

*Remark 1*: The controller design is based on the optimization of the cost function (8). In some cases, we choose $\boldsymbol{\lambda}$=**0** in minimum-phase systems to achieve the minimum tracking error provided that $M_u \geq M_y$. Additionally, static errors will increase when we increase the elements in $\boldsymbol{\lambda}$, for example, when the desired trajectory is $k^n[1,\cdots,1]_{My}^T$ ($n=1,2,\cdots$).

The essence of this kind of "model-free" method is the "model-based" method. The nature of the adaptability introduced by online identification is to reflect the real system more objectively or precisely. When we pursue good performance, especially in time-varying multivariable systems, the first and most important task is to establish the model as objectively as possible. If the established model or the estimated parameters fail to accurately reflect the actual system, the controller will not be the optimal solution to the cost function.

In other words, if the established model can accurately reflect the actual system and (16) is satisfied, the tracking performance may be guaranteed. However, in many cases, the effectiveness of the estimation may not be guaranteed. Many factors may result in the online estimated parameters hardly reflecting the true system, leading to system fragility and even output divergence. On the contrary, the controller designed using the offline estimated parameters may have enough robustness to handle modeling errors and accordingly guarantee system stability in many circumstances.

In conclusion, the objective and principle of adaptability of the controller introduced by the online estimated algorithm is to more accurately reflect the actual model of the system rather than being "model-free." This should be the guideline for this family of adaptive control methods. Therefore, the most important task is to objectively reflect the system, whether we apply the online or offline estimation method. If the online estimation method fails to meet the requirements, we may choose the offline parameters for the robustness of the controller, which may have excellent performance, even though the offline estimated parameters do not converge to the true values. Further, [26] gives a profound appreciation: "Many of the on-line schemes in current use can be thought of as sequential implementations of off-line algorithms."

[26] has put forward several questions for the guidelines. If the chosen model structure corresponds to or is able to incorporate the true system structure, i) do the estimated parameters converge to the "true" system parameters? ii) how fast does the algorithm converge? iii) how robust are the algorithms that affect various sources of errors?

Besides, we should try to incorporate the prior knowledge as much as possible into the estimation algorithm, such as structural constraints, parameter values like $\boldsymbol{\Phi}_2(k)=\boldsymbol{\Phi}_4(k)=\cdots=\mathbf{0}$ in Example 1.2, feasible ranges of parameters, etc.

To say the least, if the chosen $N$ is larger than the maximum time delay, $L_y$ and $L_u$ are large enough to encompass the controlled system model, and most importantly, we have an excellent on-line estimation method that can objectively reflect the system. The proposed MFAPC might theoretically enable the "model-free" control for many unknown nonlinear systems. It generally solves the time delay problem and is applied more widely than the current MFAC.

*Remark 2*: One implicit merit of the proposed controller is that the integrator part, $\boldsymbol{P}^T(\boldsymbol{Y}_N^*(k+1)-\boldsymbol{E}\boldsymbol{y}(k))$, is completely separated, allowing the remaining part of the controller to function as the compensator of the system model. If the controller is not properly designed, the offline model is inaccurately built, or the integral windup needs to be avoided, we may change the integrator part to

$$\begin{aligned}\boldsymbol{P}^T \Big\{ &\boldsymbol{K}_I(\boldsymbol{Y}_N^*(k+1)-\boldsymbol{E}\boldsymbol{y}(k))+ \\ &\boldsymbol{K}_P\Big[(\boldsymbol{Y}_N^*(k+1)-\boldsymbol{E}\boldsymbol{y}(k))-(\boldsymbol{Y}_N^*(k)-\boldsymbol{E}\boldsymbol{y}(k-1))\Big]\Big\}\end{aligned} \tag{24}$$

This modification allows for the adjustment of the introduced parameter matrices $\boldsymbol{K}_P$ and $\boldsymbol{K}_I$ based on the experience gained from tuning PID to change the system behavior. This step should be considered as the final procedure or last resort.

*Remark 3*: If the system is strongly nonlinear, the obtained $\boldsymbol{\phi}_L^T(k)$ $\tilde{\boldsymbol{\Psi}}_Y(k)$ $\tilde{\boldsymbol{\Psi}}_U(k)$ and $\tilde{\boldsymbol{\Psi}}_N(k)$ may change apparently from time $k$ to $k$+1, which usually causes poor system behavior. Consequently, we suggest applying the iterative MFAPC in the manner described in [23] and [24]. The controller is

$$\begin{aligned}\Delta\boldsymbol{u}(k,i)=\boldsymbol{g}^T[&\tilde{\boldsymbol{\Psi}}_N^T(k,i)\tilde{\boldsymbol{\Psi}}_N(k,i)+\boldsymbol{\lambda}(k,i)]^{-1}\tilde{\boldsymbol{\Psi}}_N^T(k,i)[(\boldsymbol{Y}_N^*(k+1) \\ &-\boldsymbol{E}\boldsymbol{y}(k,i))-\tilde{\boldsymbol{\Psi}}_Y(k,i)\Delta\boldsymbol{Y}_{Ly}(k,i)-\tilde{\boldsymbol{\Psi}}_U(k,i)\Delta\boldsymbol{U}_{Lu}(k-1,i)]\end{aligned} \tag{25}$$

, where $\tilde{\boldsymbol{\Psi}}_Y(k,i)$ , $\tilde{\boldsymbol{\Psi}}_U(k,i)$ , $\tilde{\boldsymbol{\Psi}}_N(k,i)$ , $\Delta\boldsymbol{Y}_{Ly}(k,i)$ and $\Delta\boldsymbol{U}_{Lu}(k-1,i)$ are listed in Appendix II and we choose

$$\begin{aligned}\boldsymbol{\phi}_L^T(k,i)=[\boldsymbol{\phi}_{Ly}^T(k,i),\boldsymbol{\phi}_{Lu}^T(k,i)]=[&\frac{\partial \boldsymbol{f}(\boldsymbol{\varphi}(k-1,i)}{\partial \boldsymbol{y}^T(k-1,i)},\cdots, \\ &\frac{\partial \boldsymbol{f}(\boldsymbol{\varphi}(k-1,i)}{\partial \boldsymbol{y}^T(k-n_y-1,i)},\frac{\partial \boldsymbol{f}(\boldsymbol{\varphi}(k-1,i)}{\partial \boldsymbol{u}^T(k-1,i)},\cdots,\frac{\partial \boldsymbol{f}(\boldsymbol{\varphi}(k-1,i)}{\partial \boldsymbol{u}^T(k-n_u-1,i)}]\end{aligned} \tag{26}$$

where $\boldsymbol{\phi}_L^T(k,i)$ can be calculated online by substituting the system inputs and outputs into the gradient of system function. $i$ represents the iteration count before the control inputs are sent to the system at time $k$. To save room, $\boldsymbol{\phi}_L^T(k,i)$ is shorthand for $\boldsymbol{\phi}_L^T(k+i\,|\,k)$ , $\frac{\partial \boldsymbol{f}(\boldsymbol{\varphi}(k-1,i)}{\partial \boldsymbol{y}^T(k-1,i)}$ for $\frac{\partial \boldsymbol{f}(\boldsymbol{\varphi}(k-1+i\,|\,k)}{\partial \boldsymbol{y}^T(k-1+i\,|\,k)}$ , $\Delta\boldsymbol{Y}_{Ly}(k,i)$ for $\Delta\boldsymbol{Y}_{Ly}(k+i\,|\,k)$ , and so on. For more applications, please refer to [23] and [24] which have applied the iterative MFAPC and MFAC controller with $L_y$=0 and $L_u$=1 in the robotic system. These viewpoints are the lessons from practical experiments.

## V. Conclusion

We propose a novel MFAPC method based on a prediction model derived from the corrected EDLM. The proposed method effectively addresses the time delay problem in MIMO systems and has a broader range of applications than the current MFAC. The performance analysis and the selection of the matrix $\boldsymbol{\lambda}$ are completed by analyzing the function of the closed-loop poles rather than the previous contraction mapping method which is invalid. Several simulations verify the effectiveness of the proposed method and induce the discussion about "model-free" or "model-based."

## VI. Appendix I

Proof of Theorem 1

*Proof*: Case 1: $1\le L_y\le n_y$ and $1\le L_u\le n_u$

From (1), we have

$$\begin{aligned}
&\Delta \boldsymbol{y}(k+1)=\\
&\boldsymbol{f}(\boldsymbol{y}(k),\cdots,\boldsymbol{y}(k-L_y+1),\boldsymbol{y}(k-L_y),\cdots,\boldsymbol{y}(k-n_y),\boldsymbol{u}(k),\\
&\quad\cdots,\boldsymbol{u}(k-L_u+1),\boldsymbol{u}(k-L_u),\cdots,\boldsymbol{u}(k-n_u))\\
&-\boldsymbol{f}(\boldsymbol{y}(k-1),\cdots,\boldsymbol{y}(k-L_y),\boldsymbol{y}(k-L_y),\cdots,\boldsymbol{y}(k-n_y),\boldsymbol{u}(k-1),\\
&\quad\cdots,\boldsymbol{u}(k-L_u),\boldsymbol{u}(k-L_u),\cdots,\boldsymbol{u}(k-n_u))\\
&+\boldsymbol{f}(\boldsymbol{y}(k-1),\cdots,\boldsymbol{y}(k-L_y),\boldsymbol{y}(k-L_y),\cdots,\boldsymbol{y}(k-n_y),\boldsymbol{u}(k-1),\\
&\quad\cdots,\boldsymbol{u}(k-L_u),\boldsymbol{u}(k-L_u),\cdots,\boldsymbol{u}(k-n_u))\\
&-\boldsymbol{f}(\boldsymbol{y}(k-1),\cdots,\boldsymbol{y}(k-L_y),\boldsymbol{y}(k-L_y-1),\cdots,\boldsymbol{y}(k-n_y-1),\\
&\quad\boldsymbol{u}(k-1),\cdots,\boldsymbol{u}(k-L_u),\boldsymbol{u}(k-L_u-1),\cdots,\boldsymbol{u}(k-n_u-1))
\end{aligned}\tag{27}$$

According to *Assumption* 1 and the conception of differentiability in [36], (27) becomes

$$\begin{aligned}
\Delta \boldsymbol{y}(k+1)=&\frac{\partial \boldsymbol{f}(\boldsymbol{\varphi}(k-1)}{\partial \boldsymbol{y}^T(k-1)}\Delta \boldsymbol{y}(k)+\cdots+\frac{\partial \boldsymbol{f}(\boldsymbol{\varphi}(k-1)}{\partial \boldsymbol{y}^T(k-L_y)}\Delta \boldsymbol{y}(k-L_y+1)\\
&+\frac{\partial \boldsymbol{f}(\boldsymbol{\varphi}(k-1)}{\partial \boldsymbol{u}^T(k-1)}\Delta \boldsymbol{u}(k)+\cdots+\frac{\partial \boldsymbol{f}(\boldsymbol{\varphi}(k-1)}{\partial \boldsymbol{u}^T(k-L_u)}\Delta \boldsymbol{u}(k-L_u+1)\\
&+\boldsymbol{\varepsilon}_1(k)\Delta \boldsymbol{y}(k)+\cdots+\boldsymbol{\varepsilon}_{Ly}(k)\Delta \boldsymbol{y}(k-L_y+1)\\
&+\boldsymbol{\varepsilon}_{Ly+1}(k)\Delta \boldsymbol{u}(k)+\cdots+\boldsymbol{\varepsilon}_{Ly+Lu}(k)\Delta \boldsymbol{u}(k-L_u+1)+\boldsymbol{\psi}(k)
\end{aligned}\tag{28}$$

where

$$\begin{aligned}
\boldsymbol{\psi}(k)\triangleq\ &\boldsymbol{f}(\boldsymbol{y}(k-1),\cdots,\boldsymbol{y}(k-L_y),\boldsymbol{y}(k-L_y),\cdots,\boldsymbol{y}(k-n_y),\\
&\boldsymbol{u}(k-1),\cdots,\boldsymbol{u}(k-L_u),\boldsymbol{u}(k-L_u),\cdots,\boldsymbol{u}(k-n_u))\\
&-\boldsymbol{f}(\boldsymbol{y}(k-1),\cdots,\boldsymbol{y}(k-L_y),\boldsymbol{y}(k-L_y-1),\cdots,\boldsymbol{y}(k-n_y-1),\\
&\boldsymbol{u}(k-1),\cdots,\boldsymbol{u}(k-L_u),\boldsymbol{u}(k-L_u-1),\cdots,\boldsymbol{u}(k-n_u-1))
\end{aligned}\tag{29}$$

$$\frac{\partial \boldsymbol{f}(\boldsymbol{\varphi}(k-1)}{\partial \boldsymbol{y}^T(k-i)}=\begin{bmatrix}
\frac{\partial f_1(\boldsymbol{\varphi}(k-1))}{\partial y_1(k-i)} & \frac{\partial f_1(\boldsymbol{\varphi}(k-1))}{\partial y_2(k-i)} & \cdots & \frac{\partial f_1(\boldsymbol{\varphi}(k-1))}{\partial y_{My}(k-i)}\\
\frac{\partial f_2(\boldsymbol{\varphi}(k-1))}{\partial y_1(k-i)} & \frac{\partial f_2(\boldsymbol{\varphi}(k-1))}{\partial y_2(k-i)} & \cdots & \frac{\partial f_2(\boldsymbol{\varphi}(k-1))}{\partial y_{My}(k-i)}\\
\vdots & \vdots & \vdots & \vdots\\
\frac{\partial f_{My}(\boldsymbol{\varphi}(k-1))}{\partial y_1(k-i)} & \frac{\partial f_{My}(\boldsymbol{\varphi}(k-1))}{\partial y_2(k-i)} & \cdots & \frac{\partial f_{My}(\boldsymbol{\varphi}(k-1))}{\partial y_{My}(k-i)}
\end{bmatrix}$$

$$\frac{\partial \boldsymbol{f}(\boldsymbol{\varphi}(k-1))}{\partial \boldsymbol{u}^T(k-j)}=\begin{bmatrix}
\frac{\partial f_1(\boldsymbol{\varphi}(k-1))}{\partial u_1(k-j)} & \frac{\partial f_1(\boldsymbol{\varphi}(k-1))}{\partial u_2(k-j)} & \cdots & \frac{\partial f_1(\boldsymbol{\varphi}(k-1))}{\partial u_{Mu}(k-j)}\\
\frac{\partial f_2(\boldsymbol{\varphi}(k-1))}{\partial u_1(k-i)} & \frac{\partial f_2(\boldsymbol{\varphi}(k-1))}{\partial u_2(k-i)} & \cdots & \frac{\partial f_2(\boldsymbol{\varphi}(k-1))}{\partial u_{Mu}(k-j)}\\
\vdots & \vdots & \vdots & \vdots\\
\frac{\partial f_{My}(\boldsymbol{\varphi}(k-1))}{\partial u_1(k-i)} & \frac{\partial f_{My}(\boldsymbol{\varphi}(k-1))}{\partial u_2(k-i)} & \cdots & \frac{\partial f_{My}(\boldsymbol{\varphi}(k-1))}{\partial u_{Mu}(k-j)}
\end{bmatrix}$$

, ($1\le i\le L_y$, $1\le j\le L_u$) denote the partial derivative of $\boldsymbol{f}(\cdots)$ with respect to the $i$-th vector and the $(n_y+1+j)$-th vector, respectively. And $\boldsymbol{\varepsilon}_1(k),\cdots,\boldsymbol{\varepsilon}_{Ly+Lu}(k)$ are functions that depend only on $\Delta\boldsymbol{y}(k),\cdots,\Delta\boldsymbol{y}(k-L_y+1),\Delta\boldsymbol{u}(k),\cdots,\Delta\boldsymbol{u}(k-L_u+1)$, with $(\boldsymbol{\varepsilon}_1(k),\cdots,\boldsymbol{\varepsilon}_{Ly+Lu}(k))\to(\boldsymbol{0},\cdots,\boldsymbol{0})$ when $(\Delta\boldsymbol{y}(k),\cdots,\Delta\boldsymbol{y}(k-L_y+1),\Delta\boldsymbol{u}(k),\cdots,\Delta\boldsymbol{u}(k-L_u+1))\to(\boldsymbol{0},\cdots,\boldsymbol{0})$. This means that $(\boldsymbol{\varepsilon}_1(k),\cdots,\boldsymbol{\varepsilon}_{Ly+Lu}(k))$ will be regarded as $(\boldsymbol{0},\cdots,\boldsymbol{0})$ if the control period of the system is sufficiently small.

We consider the following equation with the vector $\boldsymbol{\eta}(k)$ for each time $k$:

$$\boldsymbol{\psi}(k)=\boldsymbol{\eta}^T(k)\Delta\boldsymbol{H}(k)\tag{30}$$

Owing to $\|\Delta\boldsymbol{H}(k)\|\neq0$, (30) must have at least one solution $\boldsymbol{\eta}_0^T(k)$. Let

$$\begin{aligned}
\boldsymbol{\phi}_L^T(k)=\boldsymbol{\eta}_0^T(k)+[&\frac{\partial \boldsymbol{f}(\boldsymbol{\varphi}(k-1)}{\partial \boldsymbol{y}^T(k-1)}+\boldsymbol{\varepsilon}_1(k),\cdots,\frac{\partial \boldsymbol{f}(\boldsymbol{\varphi}(k-1)}{\partial \boldsymbol{y}^T(k-L_y)}+\boldsymbol{\varepsilon}_{Ly}(k),\\
&\frac{\partial \boldsymbol{f}(\boldsymbol{\varphi}(k-1)}{\partial \boldsymbol{u}^T(k-1)}+\boldsymbol{\varepsilon}_{Ly+1}(k),\cdots,\frac{\partial \boldsymbol{f}(\boldsymbol{\varphi}(k-1)}{\partial \boldsymbol{u}^T(k-L_u)}+\boldsymbol{\varepsilon}_{Ly+Lu}(k)]
\end{aligned}\tag{31}$$

Then (28) can be described as

$$\Delta\boldsymbol{y}(k+1)=\boldsymbol{\phi}_L^T(k)\Delta\boldsymbol{H}(k)\tag{32}$$

Case 2: $L_y=n_y+1$ and $L_u=n_u+1$

According to *Assumption* 1 and the conception of differentiability in [36], (1) becomes

$$\begin{aligned}
\Delta \boldsymbol{y}(k+1)=&\frac{\partial \boldsymbol{f}(\boldsymbol{\varphi}(k-1))}{\partial \boldsymbol{y}^T(k-1)}\Delta \boldsymbol{y}(k)+\cdots+\frac{\partial \boldsymbol{f}(\boldsymbol{\varphi}(k-1))}{\partial \boldsymbol{y}^T(k-n_y-1)}\Delta \boldsymbol{y}(k-n_y)\\
&+\frac{\partial \boldsymbol{f}(\boldsymbol{\varphi}(k-1))}{\partial \boldsymbol{u}^T(k-1)}\Delta \boldsymbol{u}(k)+\cdots+\frac{\partial \boldsymbol{f}(\boldsymbol{\varphi}(k-1))}{\partial \boldsymbol{u}^T(k-n_u-1)}\Delta \boldsymbol{u}(k-n_u)\\
&+\boldsymbol{\gamma}(k)
\end{aligned}\tag{33}$$

where

$$\begin{aligned}
\boldsymbol{\gamma}(k)=\ &\boldsymbol{\varepsilon}_1(k)\Delta\boldsymbol{y}(k)+\cdots+\boldsymbol{\varepsilon}_{Ly}(k)\Delta\boldsymbol{y}(k-n_y)\\
&+\boldsymbol{\varepsilon}_{Ly+1}(k)\Delta\boldsymbol{u}(k)+\cdots+\boldsymbol{\varepsilon}_{Ly+Lu}(k)\Delta\boldsymbol{u}(k-n_u)
\end{aligned}\tag{34}$$

Let

$$\begin{aligned}
\boldsymbol{\phi}_L^T(k)=[&\frac{\partial \boldsymbol{f}(\boldsymbol{\varphi}(k-1))}{\partial \boldsymbol{y}^T(k-1)}+\boldsymbol{\varepsilon}_1(k),\cdots,\frac{\partial \boldsymbol{f}(\boldsymbol{\varphi}(k-1))}{\partial \boldsymbol{y}^T(k-n_y-1)}+\boldsymbol{\varepsilon}_{Ly}(k),\\
&\frac{\partial \boldsymbol{f}(\boldsymbol{\varphi}(k-1))}{\partial \boldsymbol{u}^T(k-1)}+\boldsymbol{\varepsilon}_{Ly+1}(k),\cdots,\frac{\partial \boldsymbol{f}(\boldsymbol{\varphi}(k-1))}{\partial \boldsymbol{u}^T(k-n_u-1)}+\boldsymbol{\varepsilon}_{Ly+Lu}(k)]
\end{aligned}\tag{35}$$

to describe (33) as (32), with $(\boldsymbol{\varepsilon}_1(k),\cdots,\boldsymbol{\varepsilon}_{Ly+Lu}(k))\to(\boldsymbol{0},\cdots,\boldsymbol{0})$, i.e., $\boldsymbol{\phi}_L^T(k)\to[\frac{\partial \boldsymbol{f}(\boldsymbol{\varphi}(k-1))}{\partial \boldsymbol{y}^T(k-1)},\cdots,\frac{\partial \boldsymbol{f}(\boldsymbol{\varphi}(k-1))}{\partial \boldsymbol{y}^T(k-n_y-1)},\frac{\partial \boldsymbol{f}(\boldsymbol{\varphi}(k-1))}{\partial \boldsymbol{u}^T(k-1)},\cdots,\frac{\partial \boldsymbol{f}(\boldsymbol{\varphi}(k-1))}{\partial \boldsymbol{u}^T(k-n_u-1)}]$ in nonlinear systems, if $(\Delta\boldsymbol{y}(k),\cdots,\Delta\boldsymbol{y}(k-n_y),\Delta\boldsymbol{u}(k),\cdots,\Delta\boldsymbol{u}(k-n_u))\to(\boldsymbol{0},\cdots,\boldsymbol{0})$. As to linear systems, we will always have $\boldsymbol{\phi}_L^T(k)=[\frac{\partial \boldsymbol{f}(\boldsymbol{\varphi}(k-1))}{\partial \boldsymbol{y}^T(k-1)},\cdots,\frac{\partial \boldsymbol{f}(\boldsymbol{\varphi}(k-1))}{\partial \boldsymbol{y}^T(k-n_y-1)},\frac{\partial \boldsymbol{f}(\boldsymbol{\varphi}(k-1))}{\partial \boldsymbol{u}^T(k-1)},\cdots,\frac{\partial \boldsymbol{f}(\boldsymbol{\varphi}(k-1))}{\partial \boldsymbol{u}^T(k-n_u-1)}]$, no matter what $(\Delta\boldsymbol{y}(k),\cdots,\Delta\boldsymbol{y}(k-n_y),\Delta\boldsymbol{u}(k),\cdots,\Delta\boldsymbol{u}(k-n_u))$ is.

Moreover, if the function $\boldsymbol{f}(\cdots)$ has derivatives of all orders on the working points, we have (36) or (37) in accordance with the Taylor series for a function

$$\Delta y_t(k+1)=\frac{\partial \boldsymbol{f}_t(\boldsymbol{\varphi}(k-1))}{\partial \boldsymbol{H}^T(k-1)}\Delta \boldsymbol{H}(k)+\frac{1}{2!}\Delta \boldsymbol{H}^T(k)\frac{\partial^2 \boldsymbol{f}_t(\boldsymbol{\varphi}(k-1))}{\partial \boldsymbol{H}(k-1)\partial \boldsymbol{H}^T(k-1)}\Delta \boldsymbol{H}(k)+\cdots \tag{36}$$

$$\begin{aligned}\Delta y_t(k+1)=&[\sum_{p=1}^{My}\Delta y_p(k)\frac{\partial}{\partial y_p(k-1)}+\cdots+\sum_{p=1}^{My}\Delta y_p(k-n_y)\frac{\partial}{\partial y_p(k-n_y-1)}\\&+\sum_{q=1}^{Mu}\Delta u_q(k)\frac{\partial}{\partial u_q(k-1)}+\cdots+\sum_{q=1}^{Mu}\Delta u_q(k-n_u)\frac{\partial}{\partial u_q(k-n_u-1)}]f_t(\boldsymbol{\varphi}(k-1))\\&+\cdots+\frac{1}{n!}[\sum_{p=1}^{My}\Delta y_p(k)\frac{\partial}{\partial y_p(k-1)}+\cdots+\sum_{p=1}^{My}\Delta y_p(k-n_y)\frac{\partial}{\partial y_p(k-n_y-1)}\\&+\sum_{q=1}^{Mu}\Delta u_q(k)\frac{\partial}{\partial u_q(k-1)}+\cdots+\sum_{q=1}^{Mu}\Delta u_q(k-n_u)\frac{\partial}{\partial u_q(k-n_u-1)}]^n f_t(\boldsymbol{\varphi}(k-1))\\&+\cdots\\&(t=1,\cdots,M_y)\end{aligned}\tag{37}$$

, and then find one set of solutions (38), (39) for (34).

$$\boldsymbol{\varepsilon}_i(k)=\frac{1}{2!}\begin{bmatrix}\Delta \boldsymbol{y}^T(k-i+1)\dfrac{\partial^2 f_1(\boldsymbol{\varphi}(k-1))}{\partial[\boldsymbol{y}(k-i)]\partial[\boldsymbol{y}^T(k-i)]}\\ \vdots\\ \Delta \boldsymbol{y}^T(k-i+1)\dfrac{\partial^2 f_{My}(\boldsymbol{\varphi}(k-1))}{\partial[\boldsymbol{y}(k-i)]\partial[\boldsymbol{y}^T(k-i)]}\end{bmatrix}+\cdots \tag{38}$$

$$\boldsymbol{\varepsilon}_{Ly+j}(k)=\frac{1}{2!}\begin{bmatrix}\Delta \boldsymbol{u}^T(k-j+1)\dfrac{\partial^2 f_1(\boldsymbol{\varphi}(k-1))}{\partial[\boldsymbol{u}(k-j)]\partial[\boldsymbol{u}^T(k-j)]}\\ \vdots\\ \Delta \boldsymbol{u}^T(k-j+1)\dfrac{\partial^2 f_{My}(\boldsymbol{\varphi}(k-1))}{\partial[\boldsymbol{u}(k-j)]\partial[\boldsymbol{u}^T(k-j)]}\end{bmatrix}+\cdots \tag{39}$$

$i=1,\cdots,L_y$ and $j=1,\cdots,L_u$.

Case 3: $L_y>n_y+1$ and $L_u>n_u+1$

Similar to (33), (1) becomes

$$\begin{aligned}\Delta \boldsymbol{y}(k+1)=&\frac{\partial \boldsymbol{f}(\boldsymbol{\varphi}(k-1))}{\partial \boldsymbol{y}^T(k-1)}\Delta \boldsymbol{y}(k)+\cdots+\frac{\partial \boldsymbol{f}(\boldsymbol{\varphi}(k-1))}{\partial \boldsymbol{y}^T(k-n_y-1)}\Delta \boldsymbol{y}(k-n_y)\\&+\frac{\partial \boldsymbol{f}(\boldsymbol{\varphi}(k-1))}{\partial \boldsymbol{u}^T(k-1)}\Delta \boldsymbol{u}(k)+\cdots+\frac{\partial \boldsymbol{f}(\boldsymbol{\varphi}(k-1))}{\partial \boldsymbol{u}^T(k-n_u-1)}\Delta \boldsymbol{u}(k-n_u)\\&+\boldsymbol{\varepsilon}_1(k)\Delta \boldsymbol{y}(k)+\cdots+\boldsymbol{\varepsilon}_{n_y+1}(k)\Delta \boldsymbol{y}(k-n_y)\\&+\boldsymbol{\varepsilon}_{Ly+1}(k)\Delta \boldsymbol{u}(k)+\cdots+\boldsymbol{\varepsilon}_{Ly+n_u+1}(k)\Delta \boldsymbol{u}(k-n_u)\end{aligned}\tag{40}$$

Define

$$\begin{aligned}\boldsymbol{\gamma}(k)=&\boldsymbol{\varepsilon}_1(k)\Delta \boldsymbol{y}(k)+\cdots+\boldsymbol{\varepsilon}_{n_y+1}(k)\Delta \boldsymbol{y}(k-n_y)\\&+\boldsymbol{\varepsilon}_{Ly+1}(k)\Delta \boldsymbol{u}(k)+\cdots+\boldsymbol{\varepsilon}_{Ly+n_u+1}(k)\Delta \boldsymbol{u}(k-n_u)\end{aligned}\tag{41}$$

We consider the following equation with the vector $\boldsymbol{\eta}(k)$ for each time $k$:

$$\boldsymbol{\gamma}(k)=\boldsymbol{\eta}^T(k)\Delta \boldsymbol{H}(k) \tag{42}$$

Owing to $\|\Delta \boldsymbol{H}(k)\|\neq 0$, (42) must have at least one solution $\boldsymbol{\eta}_0^T(k)$. Let

$$\begin{aligned}\boldsymbol{\phi}_L^T(k)=&\boldsymbol{\eta}_0^T(k)+[\frac{\partial \boldsymbol{f}(\boldsymbol{\varphi}(k-1))}{\partial \boldsymbol{y}^T(k-1)},\cdots,\frac{\partial \boldsymbol{f}(\boldsymbol{\varphi}(k-1))}{\partial \boldsymbol{y}^T(k-n_y-1)},\boldsymbol{0},\cdots,\boldsymbol{0}\\&\frac{\partial \boldsymbol{f}(\boldsymbol{\varphi}(k-1))}{\partial \boldsymbol{u}^T(k-1)},\cdots,\frac{\partial \boldsymbol{f}(\boldsymbol{\varphi}(k-1))}{\partial \boldsymbol{u}^T(k-n_u-1)},\boldsymbol{0},\cdots,\boldsymbol{0}]^T\end{aligned}\tag{43}$$

Then (40) can be described as (32).

Case 4: $L_y\geq n_y+1$ and $1\leq L_u<n_u+1$; $0\leq L_y<n_y+1$ and $L_u\geq n_u+1$.

The proof of Case 4 is similar to the above process and is omitted.

We finished the proof of *Theorem 1*.

## VII. Appendix II

We define $\boldsymbol{\Psi}_Y(k)$, $\tilde{\boldsymbol{\Psi}}_Y(k)$, $\boldsymbol{\Psi}_U(k)$, $\tilde{\boldsymbol{\Psi}}_U(k)$, $\boldsymbol{\Psi}_N(k)$ $\tilde{\boldsymbol{\Psi}}_N(k)$, $\tilde{\boldsymbol{\Psi}}_Y(k,i)$, $\tilde{\boldsymbol{\Psi}}_U(k,i)$, $\tilde{\boldsymbol{\Psi}}_N(k,i)$ $\Delta \boldsymbol{Y}_{Ly}(k,i)$ and $\Delta \boldsymbol{U}_{Lu}(k,i-1)$ as follows:

---

$$\boldsymbol{\Psi}_N(k)_{(N\bullet My)\times(N\bullet Mu)}=\begin{bmatrix}\boldsymbol{\psi}_{11},\boldsymbol{\psi}_{12},\cdots,\boldsymbol{\psi}_{1N}\\ \boldsymbol{\psi}_{21},\boldsymbol{\psi}_{22},\cdots,\boldsymbol{\psi}_{2N}\\ \vdots\\ \boldsymbol{\psi}_{N2},\boldsymbol{\psi}_{N2},\cdots,\boldsymbol{\psi}_{NN}\end{bmatrix}$$

$$=\begin{bmatrix}\boldsymbol{\phi}_{Lu}^T(k)\boldsymbol{B} & \boldsymbol{0} & \boldsymbol{0} & \cdots & \boldsymbol{0}\\ \boldsymbol{\phi}_{Lu}^T(k+1)\boldsymbol{AB}+\boldsymbol{\phi}_{Ly}^T(k+1)\boldsymbol{D}\boldsymbol{\phi}_{Lu}^T(k)\boldsymbol{B} & \boldsymbol{\phi}_{Lu}^T(k+1)\boldsymbol{B} & \boldsymbol{0} & \cdots & \boldsymbol{0}\\ \begin{array}{c}\boldsymbol{\phi}_{Lu}^T(k+2)\boldsymbol{A}^2\boldsymbol{B}\\ +\boldsymbol{\phi}_{Ly}^T(k+2)\sum_{i=0}^{1}\boldsymbol{C}^i\boldsymbol{D}\boldsymbol{\psi}_{2-i,1}\end{array} & \begin{array}{c}\boldsymbol{\phi}_{Lu}^T(k+2)\boldsymbol{AB}\\ +\boldsymbol{\phi}_{Ly}^T(k+2)\boldsymbol{D}\boldsymbol{\phi}_{Lu}^T(k+1)\boldsymbol{B}\end{array} & \boldsymbol{\phi}_{Lu}^T(k+2)\boldsymbol{B} & \cdots & \boldsymbol{0}\\ \vdots & \vdots & \vdots & \ddots & \vdots\\ \begin{array}{c}\boldsymbol{\phi}_{Lu}^T(k+N-1)\boldsymbol{A}^{N-1}\boldsymbol{B}\\ +\boldsymbol{\phi}_{Ly}^T(k+N-1)\sum_{i=0}^{N-2}\boldsymbol{C}^i\boldsymbol{D}\boldsymbol{\psi}_{N-i-1,1}\end{array} & \begin{array}{c}\boldsymbol{\phi}_{Lu}^T(k+N-1)\boldsymbol{A}^{N-2}\boldsymbol{B}\\ +\boldsymbol{\phi}_{Ly}^T(k+N-1)\sum_{i=0}^{N-3}\boldsymbol{C}^i\boldsymbol{D}\boldsymbol{\psi}_{N-i-1,2}\end{array} & \begin{array}{c}\boldsymbol{\phi}_{Lu}^T(k+N-1)\boldsymbol{A}^{N-3}\boldsymbol{B}\\ +\boldsymbol{\phi}_{Ly}^T(k+N-1)\sum_{i=0}^{N-4}\boldsymbol{C}^i\boldsymbol{D}\boldsymbol{\psi}_{N-i-1,3}\end{array} & \cdots & \boldsymbol{\phi}_{Lu}^T(k+N-1)\boldsymbol{B}\end{bmatrix}$$

$$\tilde{\boldsymbol{\Psi}}_N(k)_{(N\cdot My)\times(N\cdot Mu)} = \boldsymbol{\Lambda}_N \boldsymbol{\Psi}_N(k)$$

$$= \begin{bmatrix} \boldsymbol{\phi}_{Lu}^T(k)\boldsymbol{B} & \mathbf{0} & \mathbf{0} & \cdots & \mathbf{0} \\ \begin{gathered}\boldsymbol{\phi}_{Lu}^T(k+1)\boldsymbol{AB}+\boldsymbol{\phi}_{Lu}^T(k)\boldsymbol{B}\\ +\boldsymbol{\phi}_{Ly}^T(k+1)\boldsymbol{D}\boldsymbol{\phi}_{Lu}^T(k)\boldsymbol{B}\end{gathered} & \boldsymbol{\phi}_{Lu}^T(k+1)\boldsymbol{B} & \mathbf{0} & \cdots & \mathbf{0} \\ \begin{gathered}\sum_{j=1}^{3}[\boldsymbol{\phi}_{Lu}^T(k+j-1)\boldsymbol{A}^{j-1}\boldsymbol{B}\\ +\boldsymbol{\phi}_{Ly}^T(k+j-1)\sum_{i=0}^{j-2}\boldsymbol{C}^i\boldsymbol{D}\boldsymbol{\psi}_{j-i-1,1}]\end{gathered} & \begin{gathered}\boldsymbol{\phi}_{Lu}^T(k+2)\boldsymbol{AB}+\boldsymbol{\phi}_{Lu}^T(k+1)\boldsymbol{B}\\ +\boldsymbol{\phi}_{Ly}^T(k+2)\boldsymbol{D}\boldsymbol{\phi}_{Lu}^T(k+1)\boldsymbol{B}\end{gathered} & \boldsymbol{\phi}_{Lu}^T(k+2)\boldsymbol{B} & \cdots & \mathbf{0} \\ \vdots & \vdots & \vdots & \ddots & \vdots \\ \begin{gathered}\sum_{j=1}^{N}[\boldsymbol{\phi}_{Lu}^T(k+j-1)\boldsymbol{A}^{j-1}\boldsymbol{B}\\ +\boldsymbol{\phi}_{Ly}^T(k+j-1)\sum_{i=0}^{j-2}\boldsymbol{C}^i\boldsymbol{D}\boldsymbol{\psi}_{j-i-1,1}]\end{gathered} & \begin{gathered}\sum_{j=2}^{N}[\boldsymbol{\phi}_{Lu}^T(k+j-1)\boldsymbol{A}^{j-2}\boldsymbol{B}\\ +\boldsymbol{\phi}_{Ly}^T(k+j-1)\sum_{i=0}^{j-3}\boldsymbol{C}^i\boldsymbol{D}\boldsymbol{\psi}_{j-i-1.2}]\end{gathered} & \begin{gathered}\sum_{j=3}^{N}[\boldsymbol{\phi}_{Lu}^T(k+j-1)\boldsymbol{A}^{j-3}\boldsymbol{B}\\ +\boldsymbol{\phi}_{Ly}^T(k+j-1)\sum_{i=0}^{j-4}\boldsymbol{C}^i\boldsymbol{D}\boldsymbol{\psi}_{j-i-1,3}]\end{gathered} & \cdots & \boldsymbol{\phi}_{Lu}^T(k+N-1)\boldsymbol{B} \end{bmatrix},$$

$$\boldsymbol{\Psi}_Y(k)=\begin{bmatrix}\boldsymbol{\varphi}_{11}\\ \boldsymbol{\varphi}_{12}\\ \boldsymbol{\varphi}_{13}\\ \boldsymbol{\varphi}_{14}\\ \vdots\\ \boldsymbol{\varphi}_{1N}\end{bmatrix}=\begin{bmatrix}\boldsymbol{\phi}_{Ly}^T(k)\\ \boldsymbol{\phi}_{Ly}^T(k+1)[\boldsymbol{C}+\boldsymbol{D}\phi_{Ly}^T(k)]\\ \boldsymbol{\phi}_{Ly}^T(k+2)\boldsymbol{C}^2+\boldsymbol{\phi}_{Ly}^T(k+2)\sum_{i=0}^{1}\boldsymbol{C}^i\boldsymbol{D}\boldsymbol{\varphi}_{12-i}\\ \vdots\\ \begin{gathered}\boldsymbol{\phi}_{Ly}^T(k+N-1)\boldsymbol{C}^{N-1}\\ +\boldsymbol{\phi}_{Ly}^T(k+N-1)\sum_{i=0}^{N-2}\boldsymbol{C}^i\boldsymbol{D}\boldsymbol{\varphi}_{1N-i-1}\end{gathered}\end{bmatrix},$$

$$\tilde{\boldsymbol{\Psi}}_Y(k)=\boldsymbol{\Lambda}_N\tilde{\boldsymbol{\Psi}}_Y(k)$$

$$=\begin{bmatrix}\boldsymbol{\phi}_{Ly}^T(k)\\ \boldsymbol{\phi}_{Ly}^T(k+1)[\boldsymbol{C}+\boldsymbol{D}\boldsymbol{\phi}_{Ly}^T(k)]\quad+\boldsymbol{\phi}_{Ly}^T(k)\\ \vdots\\ \begin{gathered}\sum_{j=1}^{N}[\boldsymbol{\phi}_{Ly}^T(k+j-1)\boldsymbol{C}^{j-1}\\ +\boldsymbol{\phi}_{Ly}^T(k+j-1)\sum_{i=0}^{j-2}\boldsymbol{C}^i\boldsymbol{D}\boldsymbol{\varphi}_{1j-i-1}]\end{gathered}\end{bmatrix},$$

$$\boldsymbol{\Psi}_U(k)=\begin{bmatrix}\boldsymbol{\varphi}_{21}^T & \boldsymbol{\varphi}_{22}^T & \boldsymbol{\varphi}_{23}^T & \cdots & \boldsymbol{\varphi}_{2N}^T\end{bmatrix}^T$$

$$=\begin{bmatrix}\boldsymbol{\phi}_{Lu}^T(k)\boldsymbol{A}\\ \boldsymbol{\phi}_{Lu}^T(k+1)\boldsymbol{A}^2+\boldsymbol{\phi}_{Ly}^T(k+1)\boldsymbol{D}\boldsymbol{\phi}_{Lu}^T(k)\boldsymbol{A}\\ \boldsymbol{\phi}_{Lu}^T(k+2)\boldsymbol{A}^3+\boldsymbol{\phi}_{Ly}^T(k+2)\sum_{i=0}^{1}\boldsymbol{C}^i\boldsymbol{D}\boldsymbol{\varphi}_{22-i}\\ \vdots\\ \boldsymbol{\phi}_{Lu}^T(k+N-1)\boldsymbol{A}^N+\boldsymbol{\phi}_{Ly}^T(k+N-1)\sum_{i=0}^{N-2}\boldsymbol{C}^i\boldsymbol{D}\boldsymbol{\varphi}_{2N-i-1}\end{bmatrix},$$

$$\tilde{\boldsymbol{\Psi}}_U(k)=\boldsymbol{\Lambda}_N\boldsymbol{\Psi}_U(k)$$

$$=\begin{bmatrix}\boldsymbol{\phi}_{Lu}^T(k)\boldsymbol{A}\\ \boldsymbol{\phi}_{Lu}^T(k+1)\boldsymbol{A}^2+\boldsymbol{\phi}_{Ly}^T(k+1)\boldsymbol{D}\boldsymbol{\phi}_{Lu}^T(k)\boldsymbol{A}+\boldsymbol{\phi}_{Lu}^T(k)\boldsymbol{A}\\ \sum_{j=1}^{3}[\boldsymbol{\phi}_{Lu}^T(k+j-1)\boldsymbol{A}^j+\boldsymbol{\phi}_{Ly}^T(k+j-1)\sum_{i=0}^{j-2}\boldsymbol{C}^i\boldsymbol{D}\boldsymbol{\varphi}_{2j-i-1}]\\ \vdots\\ \sum_{j=1}^{N}[\boldsymbol{\phi}_{Lu}^T(k+j-1)\boldsymbol{A}^j+\boldsymbol{\phi}_{Ly}^T(k+j-1)\sum_{i=0}^{j-2}\boldsymbol{C}^i\boldsymbol{D}\boldsymbol{\varphi}_{2j-i-1}]\end{bmatrix},$$

$$\tilde{\boldsymbol{\Psi}}_N(k,i)_{(N\cdot My)\times(N\cdot Mu)} =$$

$$\begin{bmatrix} \boldsymbol{\phi}_{Lu}^T(k,i)\boldsymbol{B} & \mathbf{0} & \mathbf{0} & \cdots & \mathbf{0} \\ \begin{gathered}\boldsymbol{\phi}_{Lu}^T(k+1,i)\boldsymbol{AB}+\boldsymbol{\phi}_{Lu}^T(k,i)\boldsymbol{B}\\ +\boldsymbol{\phi}_{Ly}^T(k+1,i)\boldsymbol{D}\boldsymbol{\phi}_{Lu}^T(k)\boldsymbol{B}\end{gathered} & \boldsymbol{\phi}_{Lu}^T(k+1,i)\boldsymbol{B} & \mathbf{0} & \cdots & \mathbf{0} \\ \begin{gathered}\sum_{j=1}^{3}[\boldsymbol{\phi}_{Lu}^T(k+j-1,i)\boldsymbol{A}^{j-1}\boldsymbol{B}\\ +\boldsymbol{\phi}_{Ly}^T(k+j-1,i)\sum_{m=0}^{j-2}\boldsymbol{C}^i\boldsymbol{D}\boldsymbol{\psi}_{j-m-1,1}(k,i)]\end{gathered} & \begin{gathered}\boldsymbol{\phi}_{Lu}^T(k+2,i)\boldsymbol{AB}+\boldsymbol{\phi}_{Lu}^T(k+1,i)\boldsymbol{B}\\ +\boldsymbol{\phi}_{Ly}^T(k+2,i)\boldsymbol{D}\boldsymbol{\phi}_{Lu}^T(k+1,i)\boldsymbol{B}\end{gathered} & \boldsymbol{\phi}_{Lu}^T(k+2,i)\boldsymbol{B} & \cdots & \mathbf{0} \\ \vdots & \vdots & \vdots & \ddots & \vdots \\ \begin{gathered}\sum_{j=1}^{N}[\boldsymbol{\phi}_{Lu}^T(k+j-1,i)\boldsymbol{A}^{j-1}\boldsymbol{B}\\ +\boldsymbol{\phi}_{Ly}^T(k+j-1,i)\sum_{m=0}^{j-2}\boldsymbol{C}^i\boldsymbol{D}\boldsymbol{\psi}_{j-m-1,1}(k,i)]\end{gathered} & \begin{gathered}\sum_{j=2}^{N}[\boldsymbol{\phi}_{Lu}^T(k+j-1,i)\boldsymbol{A}^{j-2}\boldsymbol{B}\\ +\boldsymbol{\phi}_{Ly}^T(k+j-1,i)\sum_{m=0}^{j-3}\boldsymbol{C}^i\boldsymbol{D}\boldsymbol{\psi}_{j-m-1.2}(k,i)]\end{gathered} & \begin{gathered}\sum_{j=3}^{N}[\boldsymbol{\phi}_{Lu}^T(k+j-1,i)\boldsymbol{A}^{j-3}\boldsymbol{B}\\ +\boldsymbol{\phi}_{Ly}^T(k+j-1,i)\sum_{m=0}^{j-4}\boldsymbol{C}^i\boldsymbol{D}\boldsymbol{\psi}_{j-m-1,3}(k,i)]\end{gathered} & \cdots & \boldsymbol{\phi}_{Lu}^T(k+N-1,i)\boldsymbol{B} \end{bmatrix}$$

$$
\tilde{\boldsymbol{\Psi}}_Y(k,i)=\begin{bmatrix} \boldsymbol{\phi}_{Ly}^T(k,i) \\ \boldsymbol{\phi}_{Ly}^T(k+1,i)[\boldsymbol{C}+\boldsymbol{D}\phi_{Ly}^T(k,i)]+\boldsymbol{\phi}_{Ly}^T(k,i) \\ \vdots \\ \sum_{j=1}^{N}[\boldsymbol{\phi}_{Ly}^T(k+j-1,i)\boldsymbol{C}^{j-1} \\ \quad +\boldsymbol{\phi}_{Ly}^T(k+j-1,i)\sum_{m=0}^{j-2}\boldsymbol{C}^i\boldsymbol{D}\boldsymbol{\varphi}_{1\,j-m-1}(k,i)] \end{bmatrix},\quad \tilde{\boldsymbol{\Psi}}_U(k,i)=
$$

$$
\begin{bmatrix} \boldsymbol{\phi}_{Lu}^T(k,i)\boldsymbol{A} \\ \boldsymbol{\phi}_{Lu}^T(k+1,i)\boldsymbol{A}^2+\boldsymbol{\phi}_{Ly}^T(k+1,i)\boldsymbol{D}\boldsymbol{\phi}_{Lu}^T(k,i)\boldsymbol{A}+\boldsymbol{\phi}_{Lu}^T(k,i)\boldsymbol{A} \\ \sum_{j=1}^{3}[\boldsymbol{\phi}_{Lu}^T(k+j-1,i)\boldsymbol{A}^j+\boldsymbol{\phi}_{Ly}^T(k+j-1,i)\sum_{m=0}^{j-2}\boldsymbol{C}^i\boldsymbol{D}\boldsymbol{\varphi}_{2\,j-m-1}(k,i)] \\ \vdots \\ \sum_{j=1}^{N}[\boldsymbol{\phi}_{Lu}^T(k+j-1,i)\boldsymbol{A}^j+\boldsymbol{\phi}_{Ly}^T(k+j-1,i)\sum_{m=0}^{j-2}\boldsymbol{C}^i\boldsymbol{D}\boldsymbol{\varphi}_{2\,j-m-1}(k,i)] \end{bmatrix},
$$

$$
\Delta\boldsymbol{Y}_{Ly}(k,i)=[\Delta\boldsymbol{y}^T(k,i),\cdots,\Delta\boldsymbol{y}^T(k-L_y+1,i)]^T,\ \Delta\boldsymbol{U}_{Lu}(k-1,i)=[\ \Delta\boldsymbol{u}^T(k-1,i),\cdots,\Delta\boldsymbol{u}^T(k-L_u,i)]^T.
$$

---